\documentclass[10pt,journal,compsoc]{IEEEtran}
\newif\ifpeerreview

\peerreviewfalse
\usepackage{booktabs}
\newcommand{\bigcell}[2]{\begin{tabular}{@{}#1@{}}#2\end{tabular}}
\usepackage{tikz}

\usepackage[nocompress]{cite}
\usepackage{url}
\usepackage{amsmath,amssymb,graphicx}
\usepackage{textcomp}

\usepackage{ragged2e}

\usepackage[switch]{lineno}
\usepackage{xcolor}
\usepackage{hyperref}
 \hypersetup{
     colorlinks=true,
     urlcolor=blue,
 }

\newcommand\copyrighttext{%
  \scriptsize \textcopyright 2020 IEEE. Personal use of this material is permitted.
  Permission from IEEE must be obtained for all other uses, in any current or future
  media, including reprinting/republishing this material for advertising or promotional
  purposes, creating new collective works, for resale or redistribution to servers or
  lists, or reuse of any copyrighted component of this work in other works.
  DOI: \href{https://doi.ieeecomputersociety.org/10.1109/TPAMI.2020.3033882}{10.1109/TPAMI.2020.3033882}.}
\newcommand\copyrightnotice{%
\begin{tikzpicture}[remember picture,overlay]
\node[anchor=south,yshift=5pt] at (current page.south) {\fbox{\parbox{\dimexpr\textwidth-\fboxsep-\fboxrule\relax}{\copyrighttext}}};
\end{tikzpicture}%
}

\title{FlatNet: Towards Photorealistic Scene Reconstruction from Lensless Measurements}

\author{Salman~S.~Khan,
        Varun~Sundar,
        Vivek~Boominathan,
        Ashok~Veeraraghavan,
        and~Kaushik~Mitra
\IEEEcompsocitemizethanks{\IEEEcompsocthanksitem S. S. Khan, V. Sundar and K. Mitra are with the Department
of Electrical Engineering, Indian Institute of Technology, Madras, TN, India, 600036.\protect\\
Email: \{sk39, varunsundar\}@smail.iitm.ac.in,kmitra@ee.iitm.ac.in

\IEEEcompsocthanksitem V. Boominathan and A. Veeraraghavan are with Rice University, Houston, TX, USA, 77005.\protect\\
Email: \{vivekb,vashok\}@rice.edu
\IEEEcompsocthanksitem S. S. Khan and V. Sundar contributed equally to this work.}
}

\begin{document}

\IEEEtitleabstractindextext{
\justify
\begin{abstract}

Lensless imaging has emerged as a potential solution towards realizing ultra-miniature cameras by eschewing the bulky lens in a traditional camera. Without a focusing lens, the lensless cameras rely on computational algorithms to recover the scenes from multiplexed measurements. However, the current iterative-optimization-based reconstruction algorithms produce noisier and perceptually poorer images. In this work, we propose a non-iterative deep learning-based reconstruction approach that results in orders of magnitude improvement in image quality for lensless reconstructions. Our approach, called \textit{FlatNet}, lays down a framework for reconstructing high-quality photorealistic images from mask-based lensless cameras, where the camera’s forward model formulation is known. FlatNet consists of two stages: (1) an inversion stage that maps the measurement into a space of intermediate reconstruction by learning parameters within the forward model formulation, and (2) a perceptual enhancement stage that improves the perceptual quality of this intermediate reconstruction. These stages are trained together in an end-to-end manner. We show high-quality reconstructions by performing extensive experiments on real and challenging scenes using two different types of lensless prototypes:  one which uses a  separable forward model and another,  which uses a  more general non-separable cropped-convolution model. Our end-to-end approach is fast, produces photorealistic reconstructions, and is easy to adopt for other mask-based lensless cameras.

\end{abstract}

}

\maketitle
\copyrightnotice


\IEEEraisesectionheading{
  \section{Introduction}\label{sec:introduction}
}

\begin{figure*}[!ht]
    \centering
    \includegraphics[width=\textwidth]{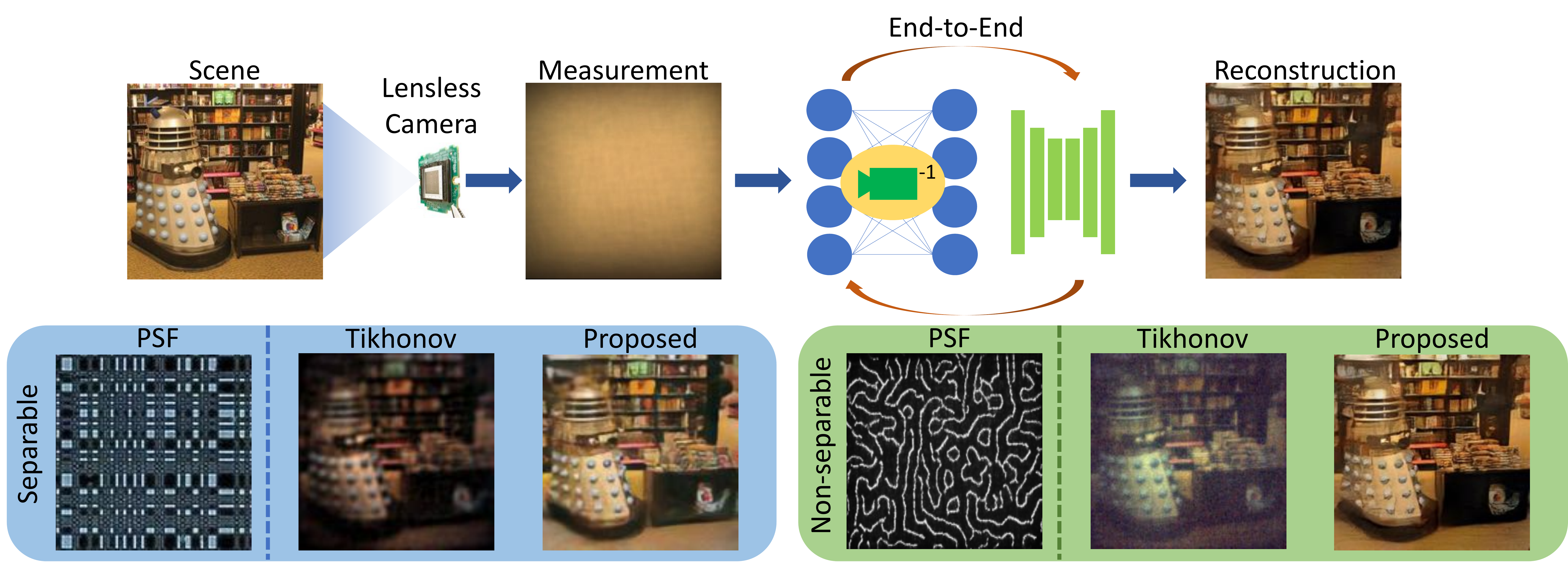}
    \caption{\textbf{Lensless imaging}. Lensless cameras require computation to recover the true scene from measurements. In this work we propose a deep learning based lensless reconstruction algorithm for both separable\cite{asif2017flatcam} and non-separable mask\cite{boominathan2020phlatcam} based lensless cameras that produce photorealistic reconstructions for real and challenging scenarios.}
    \label{fig:generic}
\end{figure*}

\IEEEPARstart{E}{merging} applications such as wearables, augmented reality, virtual reality, biometrics, and many others are driving an acute need for highly miniaturized imaging systems.
Unfortunately, current-generation cameras are based on lenses -- and these lenses typically account for more than 90\% of the cost, volume and weight of cameras.
While lenses and optics have been miniaturized by two orders of magnitude, over the last century, we are inching up against fundamental laws (diffraction limit and Lohman's scaling law\cite{lohmann1989scaling}) precluding further miniaturization.

Over the last decade, lensless imaging systems have emerged as a potential solution for light-weight, ultra-compact, inexpensive imaging.
The basic idea in lensless imaging is to replace the lens with an amplitude\cite{asif2017flatcam} or a phase mask\cite{antipa2018diffusercam,boominathan2020phlatcam}; typically placed quite close to the sensor. These lensless imaging systems provide numerous benefits over lens-based cameras. The need for a lens, which is a major contributor towards the size and weight of a camera, is eliminated. In addition, a lensless design permits a broader class of sensor geometries, allowing sensors to have more unconventional shapes (e.g.\ spherical or cylindrical) or to be physically flexible \cite{tremblay2007ultrathin}. Moreover, lensless cameras can be produced with traditional semiconductor fabrication technology and therefore exploit all of its scaling advantages - yielding low-cost, high-performance cameras \cite{boominathan2016lensless}.

Due to the absence of any focusing element, the sensor measurements recorded in a lensless imager are no longer photographs of the scene but rather highly multiplexed measurements.
Reconstruction algorithms are needed to undo the effects of this multiplexing and produce photographs of the scene being imaged. However, the design of a recovery algorithm for lensless cameras is a challenging task mainly because of the large support of the Point Spread Functions (PSFs) inherent to lensless design. In particular, the recovery algorithms face the following challenges. First, large support of PSFs result in large linear systems which makes such systems difficult to store and invert. Second, large PSFs also result in a very high degree of global multiplexing. Conventional data-driven methods like convolutional neural networks which are designed for natural images are not suited to handle this amount of multiplexing due to their limited receptive field.

Third, lensless design results in ill-conditioned systems which affect the quality of reconstruction as well as noise characteristic of such systems. The poor reconstruction quality can be observed in the Tikhonov regularized reconstructions shown in Figure \ref{fig:generic}. {Therefore, lensless cameras need robust and efficient algorithms to overcome these challenges}.\par

Keeping the above challenges in mind, we propose a feed-forward deep neural network for photorealistic lensless reconstruction, which we refer to as \textit{FlatNet}. FlatNet learns a direct mapping from lensless measurements to scene outputs. FlatNet consists of two stages: the first stage is 
a learnable inversion stage that brings the multiplexed measurements back to image space. This stage depends on the camera model. The second stage enhances this intermediate reconstruction using a fully convolutional network. 

It should be noted that the two stages are trained in an end-to-end fashion. It was shown in \cite{boominathan2020phlatcam} that separable lensless mask based lensless cameras have inferior characteristics as compared to their existing non-separable counterparts. In our previous work\cite{khan2019towards}, we had demonstrated FlatNet’s effectiveness for separable lensless model. But it cannot be trivially used for non-separable mask based lensless cameras. Here we extend the previous work to handle non-separable lensless model. In particular, we propose an efficient implementation of the learnable intermediate mapping for non-separable lensless model which is based on Fourier domain operations. {We also propose an initialization scheme for this learnable intermediate stage that doesn't require explicit PSF calibration. We show that the intermediate mapping is robust for cases where the lensless model is non-circulant}. This happens when the sensor size is smaller than the full measurement size required for deconvolution. Finally, to verify the robustness and efficiency of FlatNet, we perform extensive experiments on challenging real scenes captured using separable mask based lensless camera called FlatCam\cite{asif2017flatcam} and the non-separable mask based lensless camera called PhlatCam\cite{boominathan2020phlatcam}. To summarize, the key contributions of this paper are:

\begin{itemize}
    \item We propose an efficient implementation for the learnable intermediate stage of non-separable or general lensless model. In \cite{khan2019towards}, we had only shown this for the separable lensless model. Here we non-trivially extend it to the general lensless case. 
    \item We verify the robustness of the proposed learnable intermediate mapping for the non-separable lensless model on challenging scenarios where the lensless system does not follow a full convolutional or circulant assumption.
    \item We propose an initialization scheme for the non-separable lensless model that doesn't require explicit PSF calibration.
    \item Similar to the display and direct captured measurements collected using the separable mask FlatCam and described in our previous work\cite{khan2019towards }, we collect corresponding datasets for the non-separable mask PhlatCam\cite{boominathan2020phlatcam }.  
    \item  We also collect a dataset of unconstrained indoor lensless measurements paired with corresponding unaligned webcam images which is finally used to finetune our proposed FlatNet to robustly deal with unconstrained real-world scenes.
    \item Our method outperforms previous traditional and deep learning based lensless reconstruction methods.
\end{itemize}

\subsection{Related work}\label{related_work}

\subsubsection{Lensless imaging}\label{Lensless_imaging}
{Lensless imaging involves capturing an image of a scene without physically focusing the incoming light with a lens. It has been widely used in the past for X-ray and gamma ray imaging for astronomy \cite{dicke1968scatter,caroli1987coded}, but its use for visible spectrum applications has only recently been studied. In a lensless imaging system, the scene is captured either directly on the sensor \cite{kim2017lensless} or after being modulated by a mask element. Types of masks that have been used include phase gratings \cite{stork2013lensless}, random diffusers \cite{antipa2018diffusercam}, designed phasemasks \cite{boominathan2020phlatcam}, amplitude masks \cite{shimano2018lensless,asif2017flatcam}, compressive samplers \cite{huang2013lensless, satat2017lensless} and spatial light modulators \cite{chi2011optical,deweert2015lensless}. Replacing lens with the above masks result in multiplexed sensor capture that lacks any resemblance to the scene imaged. A recognizable image is then recovered using a computational reconstruction algorithm. In this paper, we develop a deep learning based reconstruction algorithm for both separable and non-separable mask based lensless cameras.}
\subsubsection{Image reconstruction}
{Image reconstruction is a core aspect of most computational imaging problems\cite{duarte2008single,antipa2019video,asif2017flatcam,antipa2018diffusercam,boominathan2020phlatcam}. In general, image reconstruction for computational imaging is ill-posed and requires regularization. Traditional methods for image reconstruction involve solving regularized least squares problems. Numerous regularizers based on heuristics have been developed in the past. These include the sparsity in gradient domain\cite{li2013efficient,boominathan2020phlatcam,antipa2018diffusercam}, wavelet/frequency domain sparsity\cite{reddy2011p2c2}, etc. However, these methods suffer from the fact that often the resulting cost function doesn't have a closed-form minima and an iterative approach has to be taken to solve it. Moreover, the regularizers are based on heuristics and may not be ideal for the specific task at hand.

Deep neural network have also been designed to solve image reconstruction problems in computational imaging systems. A class of deep learning based solution involves learning of regularizers or proximal mapping stage and then iteratively solving a MAP problem. Methods like \cite{dave2018solving,dave2017compressive,rick2017one} fall under this category. Another class of algorithm is designed as a feed-forward deep neural network that has either been trained in a supervised or self-supervised manner. Works on compressive image recovery\cite{kulkarni2016reconnet,mousavi2015deep,zhang2018ista}, Fourier Ptychography\cite{boominathan2018phase}, lensless recovery\cite{monakhova2019learned} fall under this category. Among these feed-forward networks, \cite{monakhova2019learned,zhang2018ista} are inspired by the physics of the imaging model and are unrolled versions of traditional optimization frameworks. Although these methods provide interpretability, the drawbacks they offer include increased computation and higher memory consumption due to large number of unrolled iterations. The proposed method and its preliminary version\cite{khan2019towards} fall under the category of physics inspired deep neural network as well. However, they don't involve any unrolling thereby avoiding large computational and memory cost.}


\section{Mask based lensless imaging}\label{lensless_description}
Mask based lensless imagers, unlike their lens-based counterparts, measure a global linear multiplexed version of the scene. This multiplexing is a function of the mask placed in front of the sensor. Mathematically, this is given as:
\begin{equation}\label{eq:gen_model}
{y} = {\Phi}{x} + {n},
\end{equation}
where  ${x}$ and $y$ are the vectorized representations of the scene and measurement respectively, ${\Phi}$ represents the generalized linear transformation, and $\textbf{n}$ is the additive noise. In general, ${\Phi}$ has a large memory footprint, and hence, storing and computing with ${\Phi}$ is computationally intractable. Reconstructing a scene with $O(N^2)$ pixels from a sensor measurement of $O(N^2)$ pixels requires ${\Phi}$ with $O(N^4)$ elements. For example, a 1-megapixel scene and a 1-megapixel sensor requires ${\Phi}$ with $\sim 10^{12}$ elements. However, by careful design of masks and using a forward model derived from physics, the computational complexity can be greatly reduced.

The modulation performed by the mask characterizes the linear matrix ${\Phi}$. By using a low-rank separable mask pattern, the huge ${\Phi}$ can be broken down into smaller matrices~\cite{asif2017flatcam,adams2017single}. Specifically, in~\cite{asif2017flatcam}, the single-separable lensless forward model reduces to:
\begin{equation}\label{eq:sep_model}
    {Y} = {\Phi}_{L}{X \Phi}_{R}^{T} + {N},
\end{equation}
where, ${\Phi}_{L}$ and ${\Phi}_{R}$ are the separable breakdown of ${\Phi}$, ${X}$ is the 2D scene irradiance, ${Y}$ is the 2D recorded measurement, and ${N}$ models additive noise.

 By adding a small enough aperture over a non-separable mask and thereby ensuring that the off-axis shifted PSF stays within the sensor,~\cite{boominathan2020phlatcam} showed that the lensless forward model can be written as a convolutional model:
\begin{equation}\label{eq:conv_model}
    {Y} = {P} \ast {X} + {N},
\end{equation}
where $P$ is PSF of the system. PSF of a lensless camera is the pattern projected by the mask on the sensor when illuminated by a single point source~\cite{boominathan2020phlatcam}. 
The PSF shifts when the point source moves laterally, and for a general scene, the sensor measurement is the weighted sum of various shifted PSFs, leading to a convolutional model. 

If the sensor isn't large enough compared to the PSF, the PSF can shift out of the sensor for an oblique angled scene point. In such a case,~\cite{antipa2018diffusercam} uses a cropped convolution model:
\begin{equation}\label{eq:cropconv}
    Y = C(P \ast X) + N,
\end{equation}
where C is the sensor cropping operation. Such a system described by Equation \ref{eq:cropconv} is no longer circulant. For a separable mask, the cropping is already incorporated in the model matrices $\Phi_L$ and $\Phi_R$.

In this work, we will be primarily focusing on two prototypes of lensless cameras, (a) FlatCam\cite{asif2017flatcam} that has a separable mask and, (b) PhlatCam\cite{boominathan2020phlatcam} that has a non-separable mask. We explore a data-driven approach that incorporates the lensless imaging models to produce photorealistic reconstructions from the above cameras.  We also explore an alternate approach to sensor cropping for PhlatCam by preprocessing the sensor measurement~\cite{reeves2005fast}.


\section{FlatNet}
\begin{figure*}[!ht]
    \begin{center}
    \includegraphics[width=\textwidth]{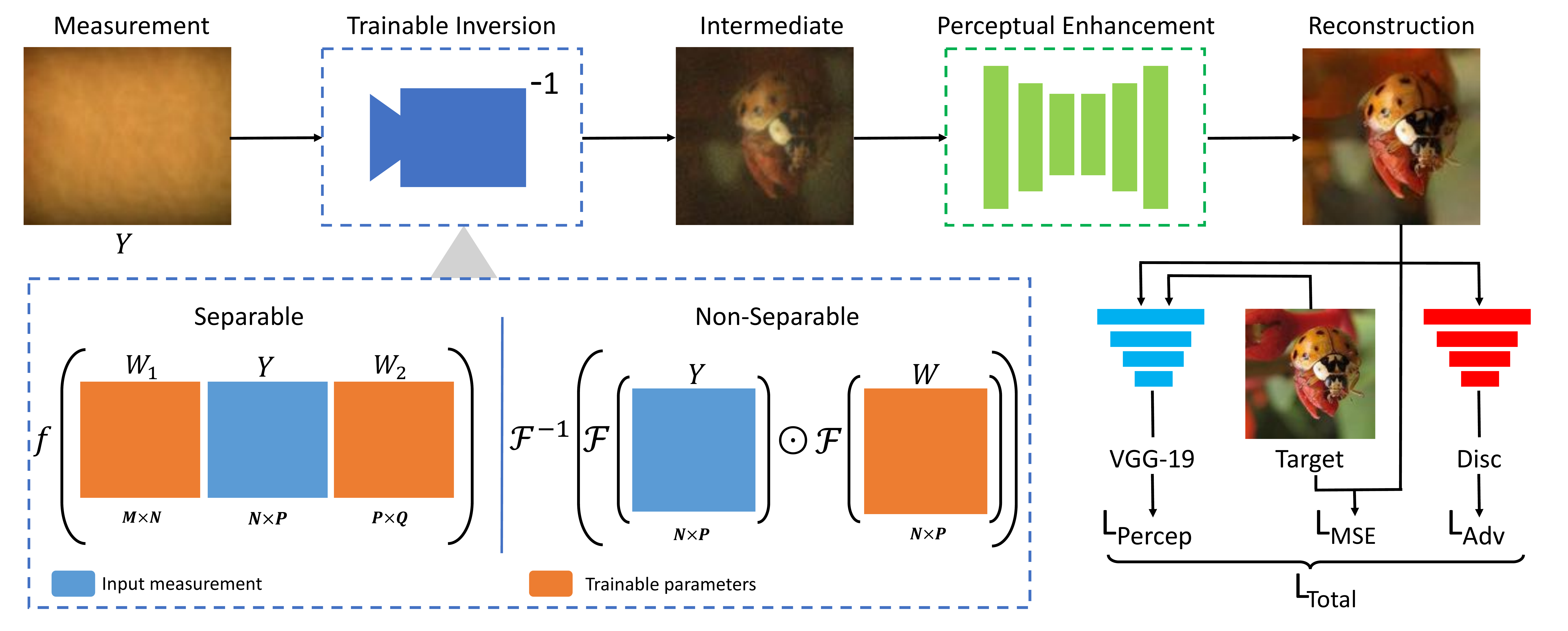}
    \caption{\textbf{Overall architecture of the FlatNet}. 
    The lensless camera measurement is first mapped into an intermediate image space using a trainable camera inversion layer. This stage is implemented separately for the separable and the non-separable case. A U-Net\cite{ronneberger2015u} then enhances the perceptual quality of the intermediate reconstruction. We use a weighted combination of three losses in training our network: a perceptual loss\cite{johnson2016perceptual} using a VGG16 network\cite{simonyan2014very}, mean-square error (MSE), and adversarial loss using a discriminator neural network \cite{goodfellow2014generative}.}
    \label{fig:main_bd}
    \end{center}
\end{figure*}
To address the challenges involved in lensless image reconstruction, we take a data-driven approach for scene recovery. We model our reconstruction framework into a two stage fully trainable deep network. This two stage network is then jointly trained in an adversarial setup. 

\textbf{Trainable camera inversion.} The first stage of FlatNet is a learnable intermediate mapping called the \textit{Trainable Camera Inversion} stage that learns to invert the lensless forward model obtaining intermediate reconstructions from globally multiplexed lensless measurements. We implement separate formulations of this trainable inversion stage for separable and non-separable lensless models exploiting the properties of the forward model for each type of these lensless systems. 

\textbf{Perceptual enhancement.} The second stage of FlatNet, called the \textit{Perceptual Enhancement} stage, is a fully convolutional network that enhances the intermediate reconstruction obtained from the trainable inversion stage giving it more photorealistic appearance. 

\subsection{Trainable camera inversion}\label{trainable_inversion}
In the first stage of our network, we learn to invert the forward operation of the lensless camera model. This allows us to obtain an intermediate representation with local structures intact. To implement this, we follow a separate approach for separable and non-separable lensless camera models. Owing to the computational simplicity of a separable model, we will first describe the implementation of the inversion stage for the separable model.
\subsubsection{Separable model}
Given the lensless model described in Equation \ref{eq:sep_model}, we learn two layers of left and right trainable matrices that act directly on 2-D measurements. This can be mathematically represented as,
\begin{equation}\label{eq:ti_sep}
    {X}_{\text{interm}} = \mathnormal{f}({W}_{1}{YW}_{2}),
\end{equation}
where ${X_{interm}}$ is the output of this stage, $\mathnormal{f}$ is a pointwise nonlinearity  (which in our case is a leaky ReLU), ${Y}$ is the input measurement, and ${W_{1}}$ and ${W_2}$ are the corresponding weight matrices for this stage. 
The dimension of the weight matrices depends on the dimension of the measurement and the scene dimension we want to recover i.e. the dimension of $W_1$ is the same as the dimension of the transpose of $\Phi_L$ while the dimension of $W_2$ is the same as the dimension of $\Phi_R$. Eventually, these matrices learn to invert the forward matrices ${\Phi_{L}}$ and ${\Phi_{R}}$. We refer to this version of FlatNet for separable lensless model as FlatNet-sep. It is important to initialize the weight matrices of this stage properly, so that the network does not get stuck in local minima. This can be done in two ways.\par
\textbf{Calibrated initialization.}
For this approach, we initialize our weight matrix ${W}_1$ with the transpose of $\Phi_{L}$ and ${W}_2$ with $\Phi_R$, akin to back-projection. These calibration matrices ($\Phi_L$ and ${\Phi}_R$) in (\ref{eq:sep_model}) are physically obtained by the method described in \cite{asif2017flatcam}. This mode of initialization leads to faster convergence while training.\par
\textbf{Uncalibrated initialization.}
Calibration of FlatCam require careful alignment with display monitor \cite{asif2017flatcam}, which can be a time consuming and inconvenient process especially for large volumes of FlatCams. Even a small error in calibration can lead to severe degradation in the performance of the reconstruction algorithm. To overcome the problems involved in calibration, we also propose a calibration-free approach by initializing the weight matrices with carefully designed pseudo-random matrices. 

Initializing with any pseudo-random matrices of appropriate size does not yield successful reconstruction. To carefully design the random initialization, we make the following two observations regarding the FlatCam forward model: the calibration matrices have a `toeplitz-like' structure and the slope of constant entries in the `toeplitz-like' structure can be approximately determined using the FlatCam geometry, in particular the distance between the mask and the sensor and the pixel pitch. As the FlatCam's geometry is known apriori, we can construct the pseudo-random `toeplitz-like' matrices with appropriate slope, and size, thereby making our approach calibration free. We discuss the generation of these pseudo-random matrices in more detail in the supplementary. The weight matrix ${W_1}$ is initialized with the adjoint of the random matrix constructed corresponding to $\Phi_L$, while the matrix ${W_2}$ is initialized with the random matrix constructed corresponding to $\Phi_R$. We observed that the training time increased slightly for this initialization in comparison to transpose initialization.\par
\subsubsection{Non-separable model}
Unlike in the separable model, it is infeasible to implement the trainable inversion stage in the non-separable model as a matrix multiplication layer owing to the extremely large dimension of ${\Phi}$. However, one can still implement it in the Fourier domain. In order to implement the inversion stage efficiently, we analyze the forward model given in Equations \ref{eq:gen_model} and \ref{eq:conv_model}.

Following the observation that the forward model is purely convolutional for an appropriate sensor dimension i.e. the forward operation is described by Equation \ref{eq:conv_model}, we model our trainable inversion stage for the non-separable case in the form of a learned inverse implemented as Hadamard product in Fourier domain. This stems from the fact that the inverse of a circulant system given by Equation \ref{eq:conv_model} is also circulant and can be diagonalized by Fourier transform.

 Mathematically, this operation is given as, 
\begin{equation}\label{eq:ti_conv}
    {X_{\text{interm}}} = \mathcal{F}^{-1}(\mathcal{F}({W})\odot \mathcal{F}({Y})),
\end{equation}
where ${X_{interm}}$ is the output of this stage and ${Y}$ is the measurement, $\mathcal{F}(.)$ and $\mathcal{F}^{-1}(.)$ are the DFT and the Inverse DFT operations, ${W}$ is the filter that is learned (akin to ${W_1}$ and ${W_2}$ in the separable model) and $\odot$ refers to Hadamard product. For a  $N\times M$ dimensional measurement, the dimension of $W$ is $N\times M$. We found that using non-linearity such as ReLU has no noticeable effect on the final output and as a result we did not include it in the non-separable model. The convolutional model of Equation \ref{eq:conv_model} would require a large sensor as the PSF's in lensless systems have large spatial dimension and in some scenarios it would be infeasible to use such a large sensor. Such a case would require the lensless model to follow Equation \ref{eq:cropconv}. Of course, we cannot accurately represent the inverse of the system described by Equation \ref{eq:cropconv} through a convolutional filter as the system is no longer circulant. As a result, one could ask if the proposed trainable inversion stage will still be valid if a smaller sensor was used? To answer this question, we show in Section \ref{smallsensor}, that with a small modification to the trainable inversion stage described in Equation \ref{eq:ti_conv}, we can handle these cropped-convolutional or non-circulant cases without significant drop in the performance. We refer to this version of FlatNet for non-separable lensless model as FlatNet-Gen.

\textbf{Calibrated initialization.} Like the separable model, initialization of ${W}$ is important for convergence of the training process. { Assuming we have a calibrated PSF and ${H}$ is the Fourier transform of this PSF, in our experiments, we initialize ${W}$ using $\mathcal{F}^{-1}(\frac{{H}^*}{K+|{H}|^2})$, i.e the regularized pseudo-inverse of the PSF or the well-known Wiener filter}. In this expression, $K$ is a regularization parameter.

\textbf{Uncalibrated initialization.} We also propose an initialization scheme that doesn't require explicit PSF calibration. { Given the mask pattern and the camera geometry, one can simulate the PSF of the lensless systems. Specifically, for PhlatCam, given the height profile of the mask, we use Fresnel propagation to simulate the PSF as described in \cite{boominathan2020phlatcam}}. This initialization scheme is particularly useful for cases where the PSF exceeds the sensor size (see Section \ref{smallsensor}). It should be noted here that this mode of initialization can be used for cases where we have access to height profile, for example in \cite{boominathan2020phlatcam}. For cases where getting a rough estimate of the height profile is not possible, for example when random diffusers are used, calibrated mode of initialization should be preferred.
\subsection{Perceptual enhancement}\label{enhancement}

Once we obtain the output of the trainable inversion stage, which is of same dimension as that of the natural image we want to recover, we use a fully convolutional network to map it to the perceptually enhanced image. Owing to its large scale success in image-to-image translation problems and its multi-resolution structure, we choose a U-Net \cite{ronneberger2015u} to map the intermediate reconstruction to the final perceptually enhanced image. We keep the kernel size fixed at 3x3 while the number of filters is gradually increased from 128 to 1024 in the encoder and then reduced back to 128 in the decoder. In the end, we map the signal back to 3 RGB channels.

For the non-seperable case, we deal with slightly larger dimensional scenes. Similar to \cite{gu2019self}, we find it useful to employ Pixel-Shuffle\cite{shi2016real} to downsample intermediate image before U-Net. By allowing U-Net to operate on a smaller spatial resolution (as a result bigger contextual area), we recover finer details for the increased image dimensions. Moreover, downsampling by Pixel-Shuffle doesn't throw away pixels and hence can be inverted exactly unlike other downsampling methods.

\subsection{Discriminator architecture}\label{discriminator}
We train FlatNet-sep and FlatNet-gen in an adversarial setup. We use a discriminator framework to classify FlatNet's output as real or fake. We find that using a discriminator network improves the perceptual quality of our reconstruction. We use 4 layers of 2-strided convolution followed by batch normalization and the swish activation function \cite{ramachandran2017searching} in our discriminator. Same discriminator architecture was used for both FlatNet-sep and FlatNet-gen.
\subsection{Loss function} \label{loss}

An appropriate loss function is required to optimize our system to provide the desired output. Pixelwise losses like mean absolute error (MAE) or mean squared error (MSE) have been successfully used to capture signal distortion. However, they fail to capture the perceptual quality of images. As our objective is to obtain high quality photorealistic reconstructions from lensless measurements, perceptual quality matters. Thus, we use a weighted combination of signal distortion and perceptual losses. The losses used for our model are given below:

\textbf{Mean squared error:} We use MSE to measure the distortion between the ground truth and the estimated output. Given the ground truth image $I_{\text{true}}$ and the estimated image $I_{\text{est}}$, this is given as:

\begin{equation}\label{eq:mse}
    \mathcal{L}_{\text{MSE}} = ||I_{\text{true}}- I_{\text{est}}||_2^2.
\end{equation}
\textbf{Perceptual loss:} To measure the semantic difference between the estimated output and the ground truth, we use the perceptual loss introduced in \cite{johnson2016perceptual}. We use a pre-trained VGG-16 \cite{simonyan2014very} model for our perceptual loss. We extract feature maps between the second convolution (after activation) and second max pool layers, and between the third convolution (after activation) and the fourth max pool layers. We call these activations $\phi_{22}$ and $\phi_{43}$, respectively. This loss is given as,
\begin{multline}\label{eq:percept}
\mathcal{L}_{\text{percept}} = \left\Vert \phi_{22}(I_{\text{true}})-\phi_{22}(I_{\text{est}})\right\Vert_{2}^{2} +\\\left\Vert \phi_{43}(I_{\text{true}})-\phi_{43}(I_{\text{est}})\right\Vert_{2}^{2}.
\end{multline}
\textbf{Adversarial loss:} Adversarial loss \cite{ledig2017photo,goodfellow2014generative} was added to further bring the distribution of the reconstructed output close to those of the real images. Given the discriminator D described in Section \ref{discriminator}, this loss is given as,
\begin{equation}\label{eq:adverse}
    \mathcal{L}_{\text{adv}} = -\log(D(I_{\text{est}})).
\end{equation}
Our discriminator, consisting of 4 layers of 2-strided convolution followed by batch normalization and ReLU activation function, classifies the generator output as real or fake.

\textbf{Total generator loss:} Our total loss for the FlatNet while training is a weighted combination of the three losses and is given as,
\begin{equation}\label{eq:tot_loss}
    \mathcal{L} = \lambda_1\mathcal{L}_{\text{MSE}} + \lambda_2\mathcal{L}_{\text{percept}} + \lambda_3\mathcal{L}_{\text{adv}}. 
\end{equation}
where, $\lambda_1$, $\lambda_2$ and $\lambda_3$ are weights assigned to each loss.

\textbf{Discriminator loss:} Given $I_{\text{est}}$, $I_{\text{true}}$ and discriminator $D$, the discriminator was trained using the following loss, 
\begin{equation}\label{eq:disc_loss}
    \mathcal{L}_{\text{disc}} = - \log(D(I_{\text{true}})) - \log(1 - D(I_{\text{est}})). 
\end{equation}

\textbf{Contextual Loss:} For finetuning FlatNet-gen on unaligned PhlatCam and webcam pairs (described in Section \ref{uncontrolled}), we use only contextual loss as proposed in\cite{mechrez2018contextual}. Denoting output image features ($\phi_{44}(I_{est})$) as $\{p_i\}_{i=1}^N $, target image features ($\phi_{44}(I_{true})$) as $\{q_j\}_{j=1}^N $ and number of pixels in each of these feature maps as $N$, contextual loss finds the nearest neighbour feature match $q =\arg \min_q \mathbb{D}(p,q_j)_{j=1}^N $ for each $p$. We then minimize the summed distance of all such feature pairs. The distance metric we adopt here  is cosine-distance, although it could also be $L_1$, $L_2$, etc. This loss term is given by:

\begin{align}\label{eq:cx}
  \mathcal{L}_{\text{contextual}} = \frac{1}{N}\sum_{i=1}^N min_{j\in [N]} \mathbb{D}(p_i, q_j) 
\end{align}

We found $\phi_{44}$ to be a suitable feature extractor based on the computational cost and sharpness of reconstruction. \par


\section{Experiments and results}\label{experiment}
In this section, we describe all our experiments. We perform all our experiments on real data. We will refer to the FlatNet for separable model as FlatNet-sep and for the non-separable model as FlatNet-gen. They will further be suffixed by -C and -UC to indicate calibrated or uncalibrated method of initialization respectively. Unless specifically mentioned, simply using FlatNet-gen or FlatNet-sep would indicate FlatNet-gen-C or FlatNet-sep-C i.e. FlatNets initialized with the calibrated method of initialization.
\begin{figure*}[!ht]
    \centering
    \includegraphics[width=\textwidth]{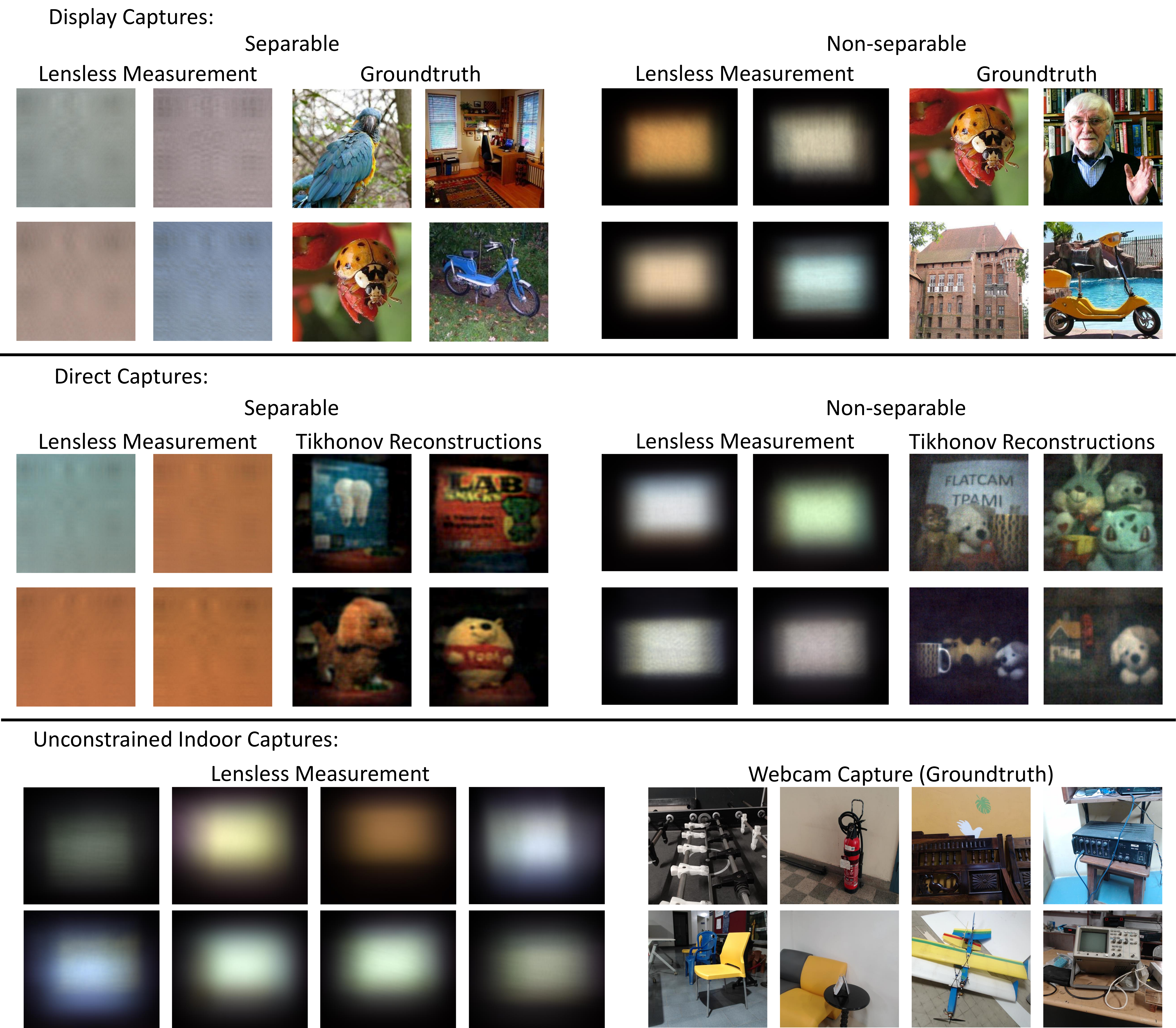}
    \caption{\textbf{Samples from our collected datasets}. All our experiments are conducted on real data captured using lensless prototypes. We collect Display Captured Dataset using both separable and non-separable prototypes to train FlatNet-sep and FlatNet-gen, respectively. We also collect Direct Captured Dataset by placing objects in front of the lensless cameras under controlled illumination. Finally, to improve the robustness of FlatNet, we collect a dataset of Unconstrained Indoor Scenes using PhlatCam and Webcam pairs.
    }
    \label{fig:dataset_samples}
\end{figure*}

\begin{figure*}[!ht]
\centering
\begin{minipage}{\textwidth}
    \centering
    \includegraphics[scale=0.44]{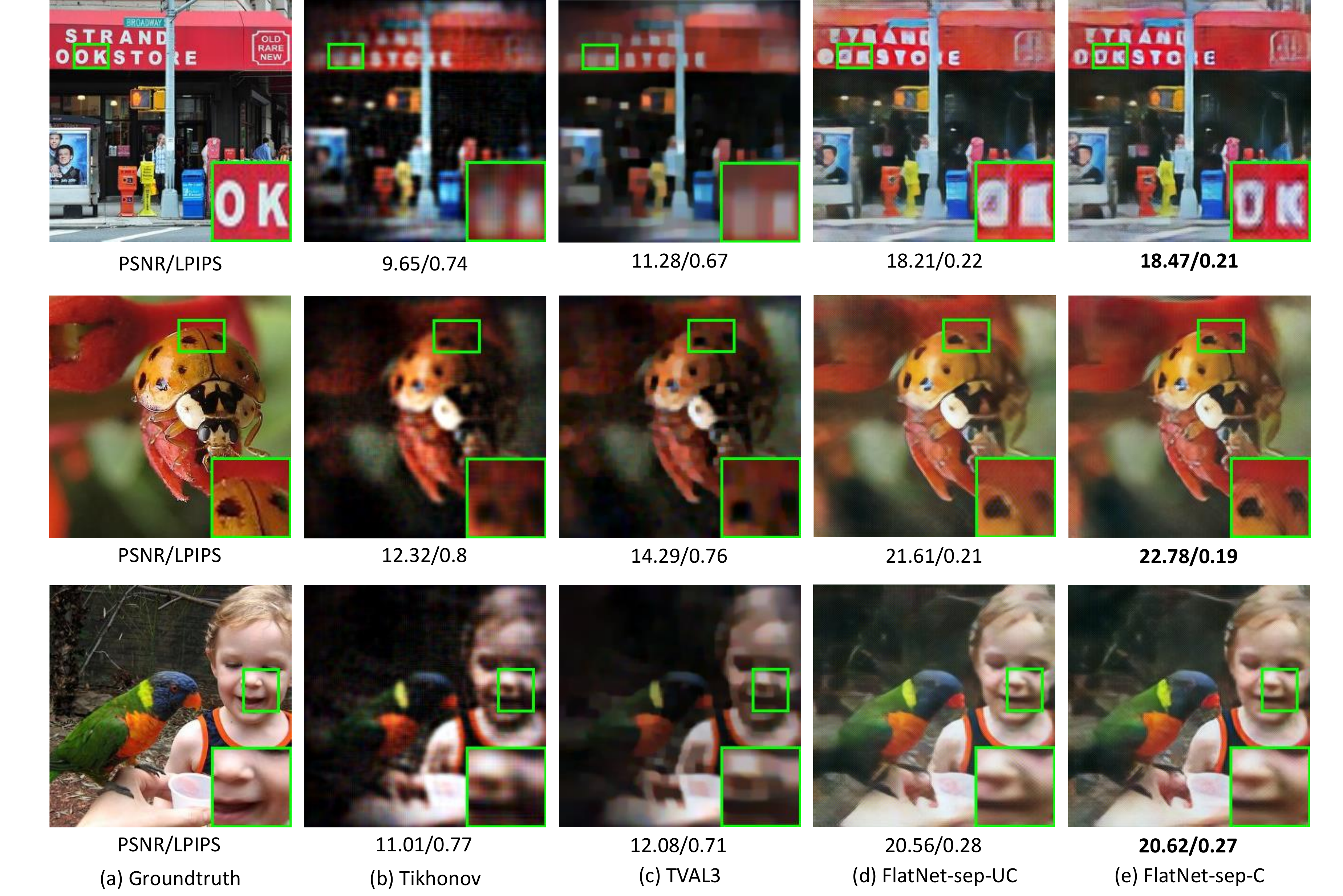}
    \caption{\textbf{Display Captured Reconstructions for FlatCam}. 
    Ground truth images are shown in a). Finer details like the text in the first image and spots on the insect in the second image are lost in b) Tikhonov regularized and c) TVAL3 reconstruction. Finer details are better preserved in FlatNet-sep for both d) uncalibrated and e) calibrated initializations.}
    \label{fig:rec_disp_amp}
\end{minipage}\\
\vspace{0.5cm}
\begin{minipage}{\textwidth}
    \centering
    \includegraphics[scale=0.44]{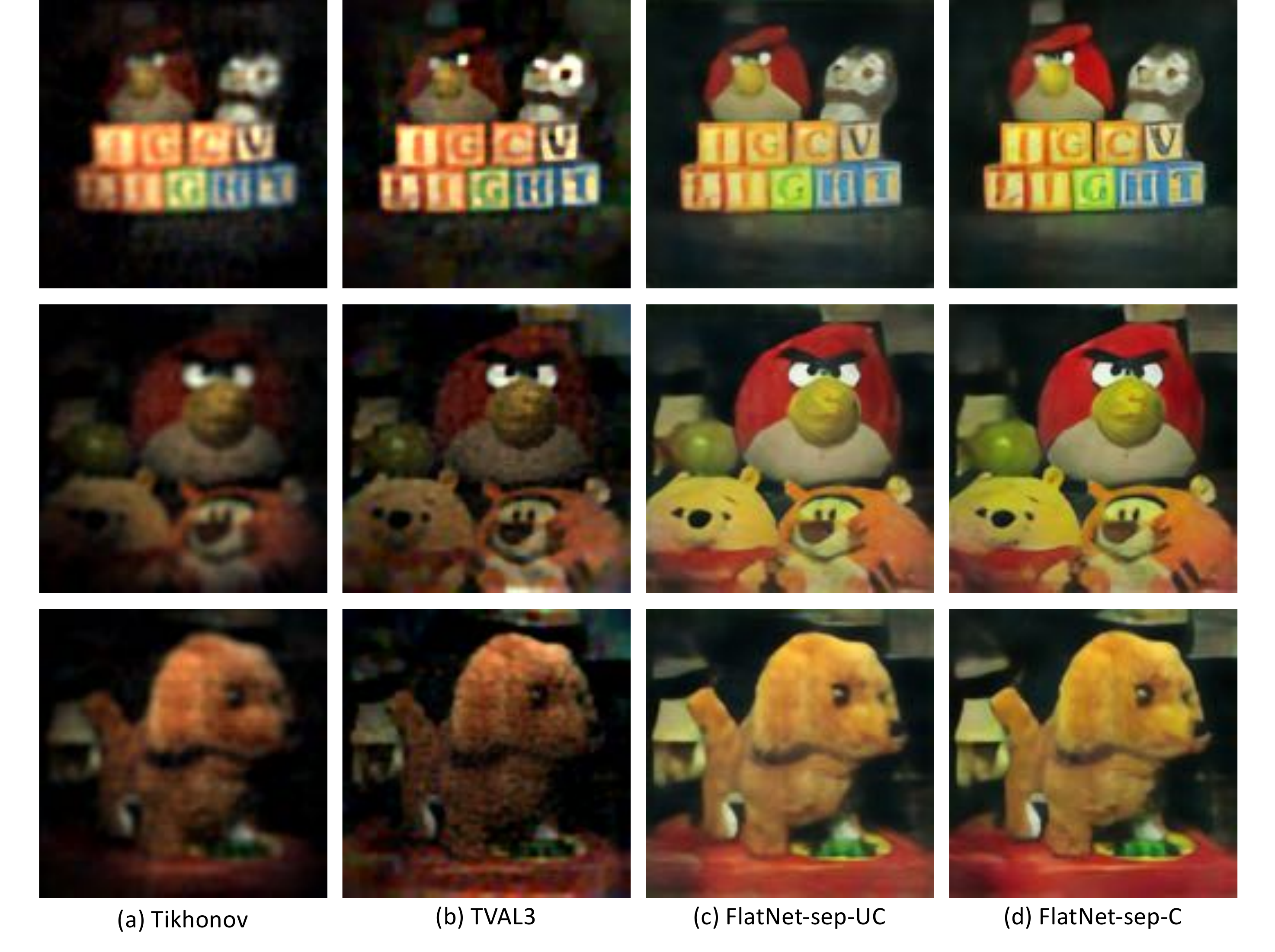}
    \caption{\textbf{Direct Captured Reconstructions for FlatCam:} a) Details in the border and darker regions are lost in the Tikhonov regularized reconstructions. b) TVAL3 reconstructs the border but is unable to restore the sharpness. The proposed end-to-end models for both c) random and d) transpose initializations produce the best reconstructions. These methods are robust to noise and does not contain any regularization parameters.}
    \label{fig:rec_dir_amp}
\end{minipage}
\end{figure*}

\subsection{Dataset}\label{dataset}
Supervised training of deep neural networks  require large scale labelled dataset. However, collecting a large scale dataset for lensless images is a challenging task. One could use the known lensless model to simulate measurements from the available natural image datasets. This, however, will sometimes fail to mimic the true imaging model due to several non-idealities. To overcome this challenge, we collect a large dataset by projecting images on monitors and capturing this projection using lensless cameras.  This not only takes care of the true imaging model for lensless camera, it also helps us collect a labelled dataset for lensless images. We follow the same dataset collection procedure for both FlatCam\cite{asif2017flatcam} and PhlatCam\cite{boominathan2020phlatcam}. For our work, we use a subset of ILSVRC 2012\cite{ILSVRC15}. Specifically, we used 10 random images from each class as our ground truth. Of the 1000 classes, we kept 990 classes for training and the rest for testing. So in total, we used 9900 images for training and 100 images for testing. Before capturing the dataset, we resize the images displayed on monitor so as to cover the entire field of view (FoV) of camera. We call this dataset the Display Captured Dataset. For this dataset, the ground truth images are the ones that were projected on the monitor screen. The monitor was kept beyond the hyperfocal distance of the cameras to avoid the variation of the PSF with depth. The hyperfocal distance for the FlatCam prototype is around a foot and for the PhlatCam prototype is around 16 inches. To test the FlatNet on real scenes, we also capture measurements of objects placed directly in front of the camera. Using FlatCam we collect 15 such measurements while using PhlatCam we collect 20 such measurements. We call this dataset Direct Captured Dataset. This dataset doesn't have corresponding ground truths for the measurements. To demonstrate the effectiveness of FlatNet-gen on unconstrained indoor scenarios, we collect a dataset of unaligned PhlatCam and webcam captures using the setup described in Figure \ref{fig:finetune}. This dataset consists of 475 training samples and 25 test samples. We call this dataset the Unconstrained Indoor Dataset. Samples from our datasets can be seen in Figure \ref{fig:dataset_samples}. Code and dataset have been made available\footnote{Link: https://github.com/siddiquesalman/flatnet}.

\begin{figure*}[!ht]
    \centering
    \includegraphics[scale=0.3]{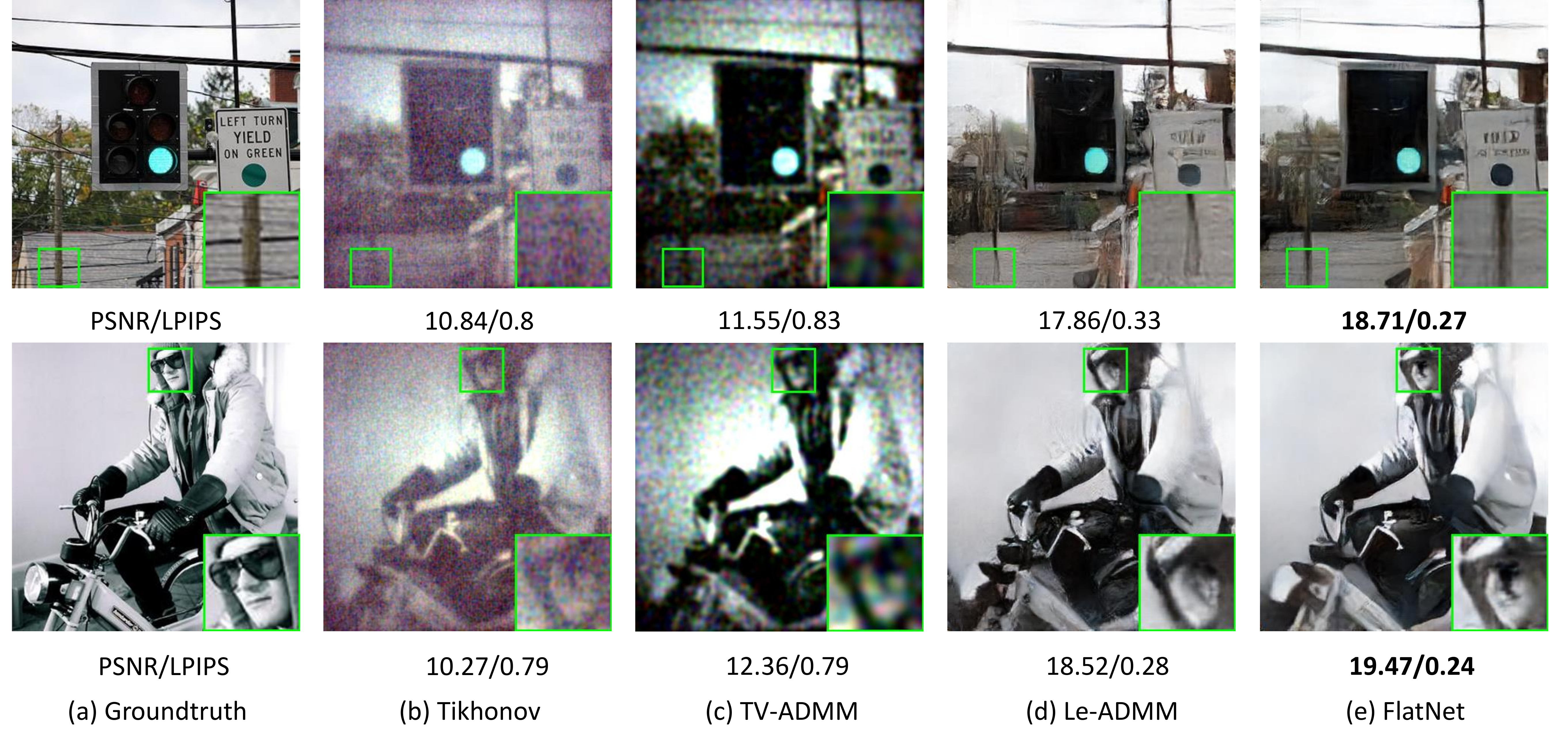}
    \caption{\textbf{Display Captured Reconstructions for PhlatCam}. While the learning based methods clearly outperform traditional methods like Tikhonov and TV-based ADMM, FlatNet-gen has superior performance in terms of reconstructing finer details.}
    \label{fig:rec_disp_phase}
\end{figure*}
\begin{figure*}[!ht]
    \centering
    \includegraphics[scale=0.3]{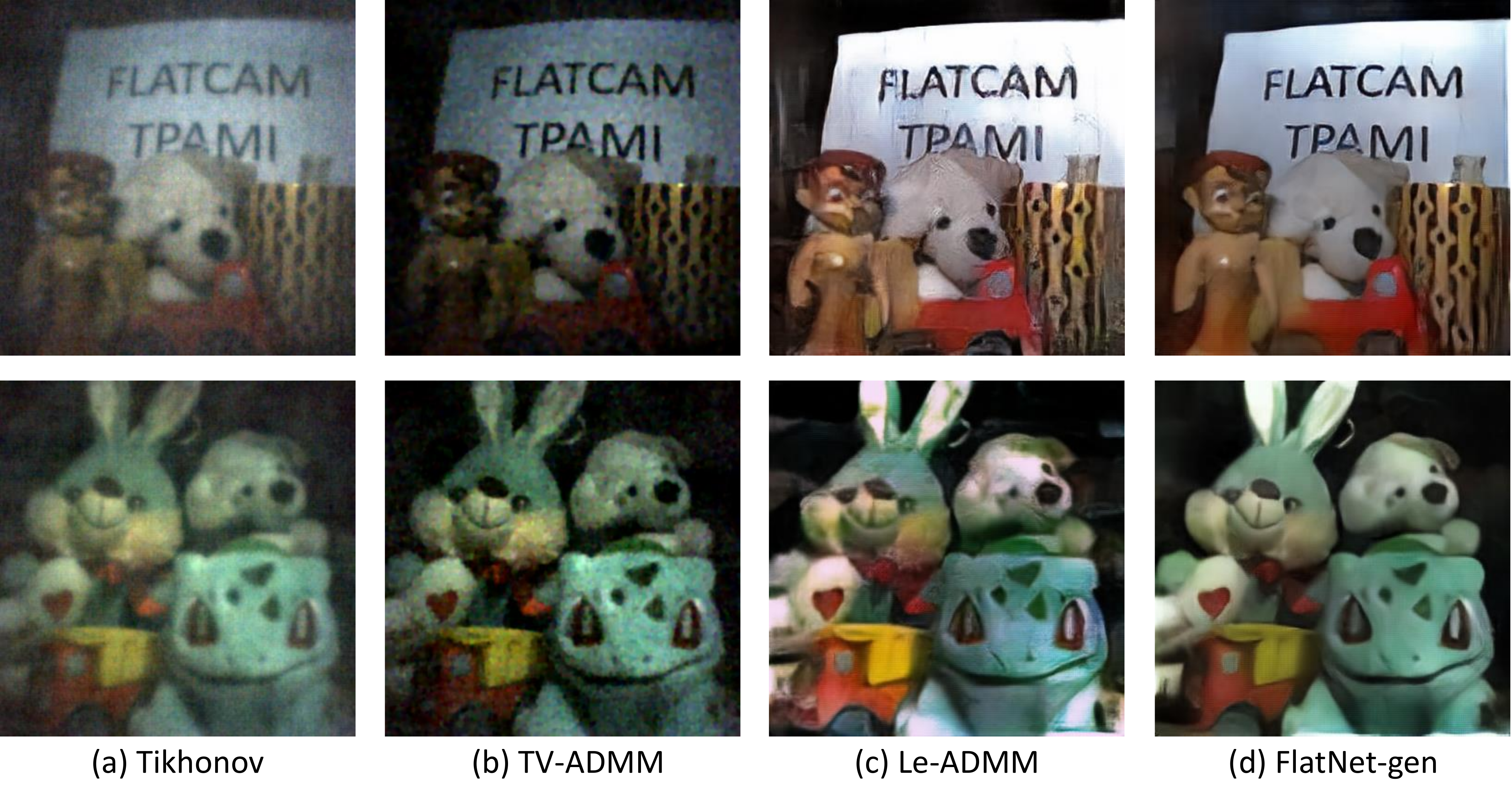}
    \caption{\textbf{Direct Captured Reconstructions for PhlatCam}. FlatNet-gen has fewer artifacts while Le-ADMM suffers from blurry reconstructions and hallucinated artifacts.}
    \label{fig:rec_dir_phase}
\end{figure*}
\subsection{Implementation details}\label{implementation}

The FlatCam prototype uses a Point Grey Flea3 camera with 1.3MP e2v EV76C560 CMOS sensor and a pixel size of 5.3 $\mu$m. All the ground truth images were resized to $256\times 256$ as the FlatCam is calibrated to produce $256\times 256$ output images. This ensures that there is no misalignment among the input and ground truth pairs. We directly used the Bayer measurements, split into 4 channels (R,Gr,Gb,B), as our input to the network and convert them into 3 channel RGB within the network. FlatCam measurements of dimension $512\times 640\times 4$ in batches of 4 were used as inputs for training. A smaller batch size was used due to memory constraints. We set $\lambda_1$ as 1, $\lambda_2$ to be 1.2 and $\lambda_3$ to be 0.6. For transpose initialization, we trained our model for 45K iterations while for random initialization, we trained it for 60K iterations. The Adam\cite{kingma2014adam} optimizer was used for all models. We started with a learning rate of $10^{-4}$ and gradually reduced it by half every 5000 iterations. 
The PhlatCam prototype used is a Basler Ace4024-29uc with 12.2MP Sony IMX226 sensor with a pixel size of 1.85$\mu$m. All the ground truth images were resized to $384\times 384$ which is equal to the FoV of the prototype. We directly used the Bayer measurements, split into 4 channels (R,Gr,Gb,B), as our input to the network and convert them into 3 channel RGB within the network. We used the same set of $\lambda_{i}$'s as that for FlatNet-sep. The full measurements used were of dimension $1280\times 1408\times 4$. 
{ For the small sensor experiments of Section \ref{smallsensor}, we use measurements of dimension $608\times 864\times 4$}. 
\subsection{Comparison with other approaches}\label{comparison}

\subsubsection{Separable lensless model}\label{separable}
In this subsection, we show results for the amplitude mask FlatCam that follows a separable model.\par
We compare FlatNet-sep with the closed form Tikhonov reconstruction described in \cite{asif2017flatcam} and a total variation based reconstruction implemented using TVAL3\cite{li2013efficient}.\par

\begin{table}[ht]
\centering
\caption{\textbf{Average Metrics on Display Captured FlatCam measurements}. FlatNet-sep with transpose initialization (FlatNet-sep-C) gives the best result. Comparable performance of FlatNet-sep-UC indicates that our approach can be used for situations where careful calibration isn't possible.}
\resizebox{\columnwidth}{!}{
\begin{tabular}{ |c||c|c|c|c| }
\hline
\textbf{Method}&\textbf{PSNR (in dB)} &\textbf{SSIM}& \textbf{LPIPS}&\bigcell{c}{\textbf{Inference} \\ \textbf{Time (in sec)}}\\
\hline
Tikhonov   &  10.95 &0.33  & 0.795 & 0.03\\
\hline
TVAL3&  11.81  & 0.36&0.752&45.28\\
\hline
{FlatNet-sep-UC}    &19.06 &0.62&0.274&0.006\\
\hline
\textbf{FlatNet-sep-C}    &\textbf{19.62} &\textbf{0.64}&\textbf{0.256}&\textbf{0.006}\\
\hline
\end{tabular}}
\label{tab:table1}
\end{table}

\textbf{Qualitative discussion.}
In Figure \ref{fig:rec_disp_amp}, we compare our methods, FlatNet-sep-UC with uncalibrated initialization and FlatNet-sep-C with calibrated initialization, with traditional methods, Tikhonov and TVAL3.
As can be observed from the reconstructions, the Tikhonov regularized reconstructions are prone to severe vignetting effects which is somewhat reduced in the TVAL3 results. Inset images in Figure \ref{fig:rec_disp_amp} show the preservation of finer details in our approach. Figure \ref{fig:rec_dir_amp} shows the performance of the various methods for direct captured measurements. Tikhonov regularization has a tendency to suppress low signal values and as a result has difficulty restoring the poorly illuminated background for most of the scenes in Figure \ref{fig:rec_dir_amp}. The performance of TVAL3\cite{li2013efficient} is also similar. FlatNet-sep, on the other hand, produces higher quality photorealistic reconstruction. Note that our uncalibrated model FlatNet-sep-UC gives similar performance to that of the calibrated model FlatNet-sep-C. Thus, our method does not require explicit calibration unlike the rest of the approaches.\par
\textbf{Quantitative discussion.}
We present the quantitative performance of FlatNet for separable mask FlatCam in Table \ref{tab:table1}. For evaluation, we use PSNR, SSIM and the recently proposed LPIPS\cite{zhang2018unreasonable}. Higher PSNR and SSIM score indicate better performance while lower LPIPS indicates better perceptual quality. It can be clearly seen that our approach using transpose initialization (FlatNet-sep-C) outperforms all the other reconstruction techniques for FlatCam. The next best approach is the FlatNet-sep using random initialization (FlatNet-sep-UC), which unlike other methods, is a calibration-free technique. We also compare the inference time for various approaches in the same table. The Tikhonov and TVAL3\cite{li2013efficient} regularized reconstructions are computed on Intel Core i7 CPU with 16 GB RAM while the rest of the approaches are evaluated on Nvidia GTX 1080 Ti GPU.

\subsubsection{Non-separable lensless model}\label{nonsep}
For experiments on the non-separable model, we compare FlatNet-gen with traditional and learning based approaches. We describe these approaches below.\par
\textbf{Traditional approaches.} In traditional method, we compare FlatNet-gen with traditional Tikhonov regularized reconstruction implemented in Fourier domain (as Wiener restoration filter) and total variation regularized reconstruction implemented using ADMM\cite{antipa2018diffusercam}\par
\textbf{Learning based approaches.} For learning based approach, we use the unrolled deep network described in \cite{monakhova2019learned}. However, for fairness, we use the five stage unrolled ADMM followed by our perceptual enhancement stage. \par

\textbf{Qualitative discussion.}
Figure \ref{fig:rec_disp_phase} shows the display captured reconstruction for PhlatCam. We can clearly see higher quality reconstruction for FlatNet-gen in comparison to traditional Tikhonov regularized reconstruction or Wiener deconvolution and ADMM based method. It also results in better quality reconstruction than the Le-ADMM model. This trend in performance is also observed in the direct captured reconstructions in Figure \ref{fig:rec_dir_phase}. It should also be noted that Le-ADMM, despite having fewer parameters, is extremely memory and computation Wintensive due to the large number of intermediates/primal and dual variables calculated at each stage of the unrolled ADMM. It is due to this significant increment in memory consumption, that it becomes infeasible to implement this model on the captured PhlatCam measurements without downsampling. In our comparison, we downsample the measurements by a factor of 4 (similar to \cite{monakhova2019learned}) before passing them through the Le-ADMM network. Unless explicitly mentioned, we will refer to this downsampled Le-ADMM model as Le-ADMM. Downsampling operation leads to compromise in the reconstruction resolution resulting in the lack of sharpness observed in the final reconstruction. On the other hand, the FlatNet-gen has significantly lower memory requirement that doesn't require any downsampling pre-processing thereby preventing any loss of sharpness or resolution. We also provide comparison for FlatNet-gen initialized with uncalibrated PSF in the supplementary material. We call this model FlatNet-gen-UC.\par

\begin{table}[!t]
\centering
\caption{\textbf{Average Metrics on Display Captured PhlatCam measurements}. FlatNet-gen produces higher quality results without compromising on the inference time for both the real PSF case (FlatNet-gen-C) and the simulated PSF case (FlatNet-gen-UC). Le-ADMM shows larger difference in quality between the real and simulated PSF cases owing to its stronger dependence on the PSF.}
\resizebox{\columnwidth}{!}{
\begin{tabular}{ |c||c|c|c|c| }
\hline
\textbf{Method}&\textbf{PSNR (in dB)} &\textbf{SSIM}& \textbf{LPIPS}&\bigcell{c}{\textbf{Inference} \\ \textbf{Time (in sec)}} \\
\hline
Tikhonov  &  12.67 &0.25  & 0.758 & 0.03\\
\hline
TV-ADMM&  13.51  & 0.26 &0.755 &180\\
\hline
Le-ADMM-UC   &  18.35 & 0.49   & 0.407&0.08\\
\hline
Le-ADMM-C   &  20.29 & 0.51   & 0.333 &0.08\\
\hline
FlatNet-gen-UC    & 20.53 & 0.54 & 0.318 & 0.03\\
\hline
\textbf{FlatNet-gen-C}    &\textbf{20.94} &\textbf{0.55}&\textbf{0.296}&\textbf{0.03}\\
\hline
\end{tabular}}
\label{tab:table2}
\end{table}

\textbf{Quantitative discussion.}
The quantitative results are provided in Table \ref{tab:table2}. { Along with the uncalibrated FlatNet-gen model, we also provide the performance of uncalibrated version of  Le-ADMM in this table. It is referred to as Le-ADMM-UC.}  The consistency with visual results is maintained in the quantitative metrics. It can be clearly seen that FlatNet-gen outperforms all other methods quantitatively. FlatNet-gen-UC performs almost at par with FlatNet-gen-C and outperforms Le-ADMM-UC. It should be noted that the difference between FlatNet-gen-C and FlatNet-gen-UC is smaller as compared to Le-ADMM-C and Le-ADMM-UC. This is primarily due to the stronger dependence of Le-ADMM on the true PSF while FlatNet-gen requires the knowledge of PSF only for better initialization and learns to converge to a better inverse after training. We also provide the runtime for the methods compared. For Wiener and TV-based ADMM, we report the speed on CPU while for others we report the speed for a forward pass in GPU.

Assuming the true measurement is of dimension $1280\times 1408$, we additionally compare FlatNet-gen's trainable inversion stage with the unrolled ADMM block of Le-ADMM (without the U-Net) in terms of memory and computation in Table \ref{tab:table4}. We provide the memory consumption (in Megabytes, computed on Nvidia GTX 1080 Ti GPU) and computations (in FLOPs, computed theoretically) required to process one image using the two methods. We unroll the ADMM for 5 iterations. In the table, Le-ADMM-Full refers to the unrolled ADMM without any downsampling  while Le-ADMM-Downsampled refers to the case where the PSF and the scene were downsampled by a factor of 4. { It can be observed that a full resolution Le-ADMM requires significant amount of memory which would have negative implications if deployment is considered. Moreover, appended with dense CNNs like U-Net, Le-ADMM-Full is difficult to implement on a conventional GPU, thereby necessitating the downsampling of the measurements which in turn leads to the degradation of the reconstruction quality}. One should also note the amount of computations performed in the unrolled ADMM block for the particular dimensions of PSF and scene. Due to a series of intermediate estimates that depend on Fourier and Inverse Fourier transforms, this computation blows up for Le-ADMM-Full. FlatNet-gen provides a better trade-off for resolution, and memory and computational requirements which is essential for lensless systems which, by design, suffer from poor reconstruction resolution.

\begin{table}[!t]
\centering
\caption{\textbf{Memory and FLOP comparison}. Comparison of memory consumption and FLOPs for five unrolled iterations of the ADMM block in Le-ADMM (full and 4X downsampled versions) and the trainable inversion stage of our proposed FlatNet-gen. 
}
\begin{tabular}{ |c||c|c|c| }
\hline
\textbf{Method}&\bigcell{c}{\textbf{Memory} \\ \textbf{(in MB)}}  &\bigcell{c}{\textbf{Computation} \\ \textbf{(in MFLOP)}}\\
\hline
Le-ADMM-Full  &  6300& 1290\\
\hline
\bigcell{c}{Le-ADMM-\\Downsampled}  &  1000 & 65\\ 
\hline
\textbf{FlatNet-gen}    &\textbf{990}  &\textbf{53}\\ 
\hline
\end{tabular}
\label{tab:table4}
\end{table}

\subsection{Further analysis}
\subsubsection{Effect of learning the inversion stage}\label{naive_comp}

\begin{figure}[!t]
\centering
    \includegraphics[scale=0.23]{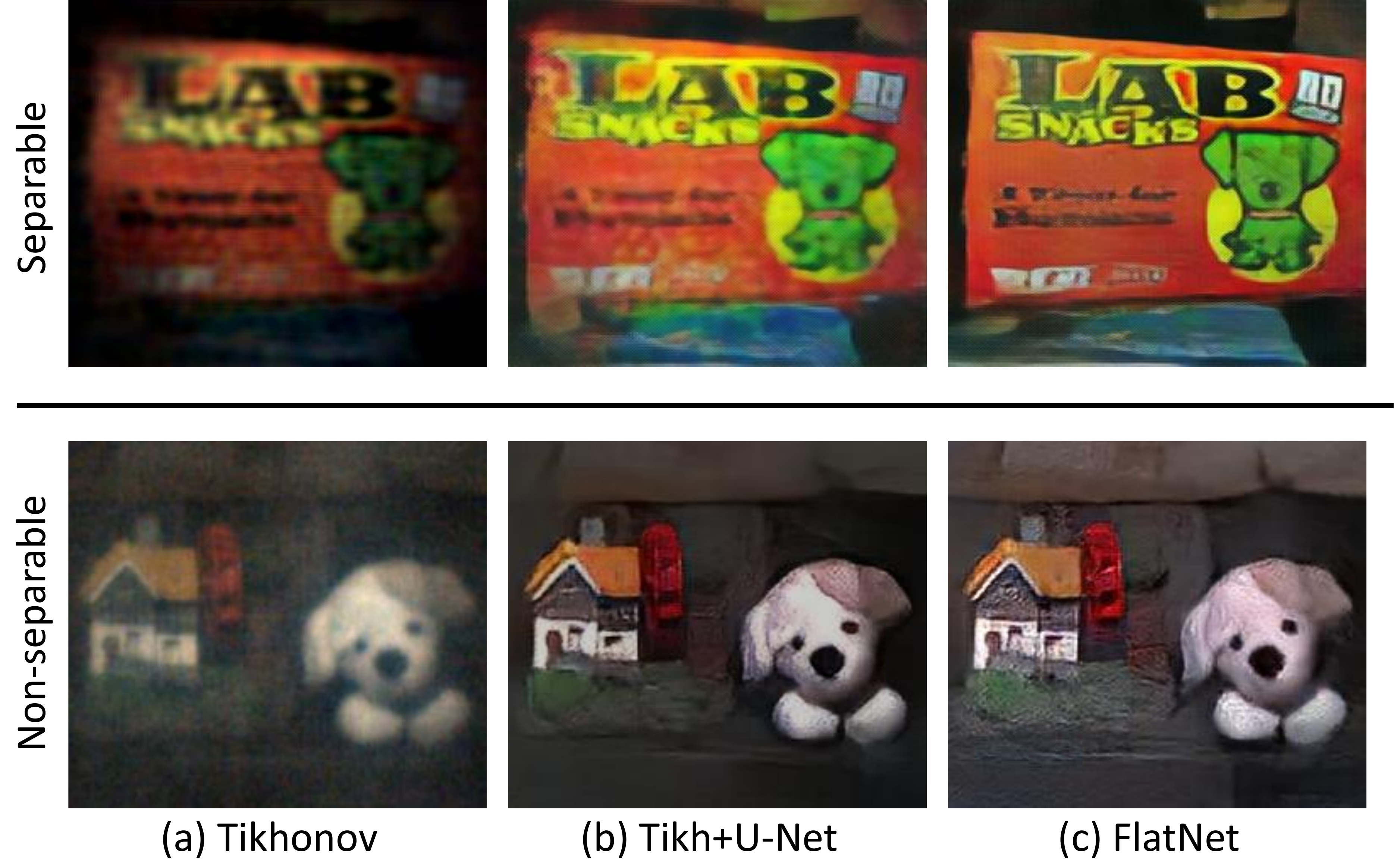}
    \begin{center}
    \caption{\textbf{Comparison of FlatNet with Tikh+U-Net}. {Top row shows the comparison of FlatNet-sep with Tikh+U-Net while the bottom row shows the comparison of FlatNet-gen with Tikh+U-Net. FlatNet provides sharper and more photorealistic reconstructions compared to Tikh+U-Net for both separable and non-separable models}.}
    \label{fig:naive_sep_gen}
    \end{center}
\end{figure}

In this section, we highlight the importance of the end-to-end learning strategy of FlatNet.{ We compare FlatNet with a network with just the perceptual enhancement block. We train this network with Tikhonov regularized reconstructions. For training this network, we use the same loss as defined in Equation \ref{eq:tot_loss}}. We call this method Tikh+U-Net. We implement this approach for both separable and non-separable lensless models. Top row of Figure \ref{fig:naive_sep_gen} compares the reconstruction quality of FlatNet-sep with Tikh+U-Net. We can easily observe the improved quality of reconstruction obtained from FlatNet-sep compared to Tikh+U-Net. Tikh+U-Net suffers from blurrier reconstructions with amplified artifacts. We also compare the performance of FlatNet-gen with its corresponding Tikh+U-Net in the bottom row of Figure \ref{fig:naive_sep_gen}. FlatNet-gen provides sharper reconstructions over Tikh+U-Net.

Table \ref{tab:naive} provides a quantitative flavor to the above analysis. We can see that FlatNet outperforms Tikh+U-Net for both separable and non-separable models in terms of PSNR and LPIPS. 

One may notice that the difference between FlatNet-gen and Tikh+U-Net is not as significant as between FlatNet-sep and its corresponding Tikh+U-Net. This is due to the higher quality of Tikhonov reconstruction in the case of PhlatCam compared to FlatCam\cite{boominathan2020phlatcam}. However, one should note that Tikh+U-Net is strictly based on convolutional assumption for the forward model, and performs poorly when this assumption is violated as will be verified in Section \ref{smallsensor}.

\begin{table}[!t]
\centering
\caption{\textbf{Comparison of FlatNet with Tikh+U-Net}. The top half compares FlatNet-sep with Tikh+U-Net for separable lensless model while the bottom half compares FlatNet-gen with the corresponding Tikh+U-Net. FlatNet outperforms Tikh+U-Net for both separable and non-separable models because it learns an end-to-end mapping.}
\begin{tabular}{|c|cc|cc}
 \hline
 \textbf{Methods} & \textbf{PSNR (in dB)} & \textbf{LPIPS}\\ \hline
\textbf{Separable Model} &&\\ \hline
Tikh+U-Net & 18.90 & 0.322\\
\textbf{FlatNet} & \textbf{19.62} & \textbf{0.256}\\
\hline
\textbf{Non-separable Model} &&\\ \hline
Tikh+U-Net & 20.60 & 0.298\\ 
\textbf{FlatNet} & \textbf{20.94} & \textbf{0.296}\\
 \hline
\end{tabular}
\label{tab:naive}
\end{table}

\subsubsection{Performance on cropped measurements}\label{smallsensor}

As we have already seen in Section~\ref{lensless_description}, the forward operation in a mask-based lensless camera is no longer convolutional if the size of the sensor is small compared to the true measurement size i.e. the forward model is given by Equation~\ref{eq:cropconv}. This coupled with large PSFs, makes lensless reconstruction challenging for traditional reconstruction approaches which rely on the circulant or convolutional assumptions (e.g. Wiener deconvolution). This naturally leads to a question: Will the proposed trainable inversion layer of FlatNet-gen, which is based on learned Fourier domain inversion, be robust against cases where the deviation from the circulant assumption is significant? In other words, will FlatNet-gen be able to deal with measurements from which a significant amount of pixels have been thrown away due to the finite sensor size and fully open aperture? In this section, we show that we can deal with the small sensor size case without losing much in terms of reconstruction quality and perform better than Le-ADMM which explicitly tries to deal with the cropped out pixels. For our experiments, we take a central crop of size $608\times 864$ from our 7MP full sensor measurement. Effectively, this can be thought as using a 2MP sensor instead of the 7MP sensor.

\begin{figure}[!t]
\centering
    \includegraphics[scale=0.19]{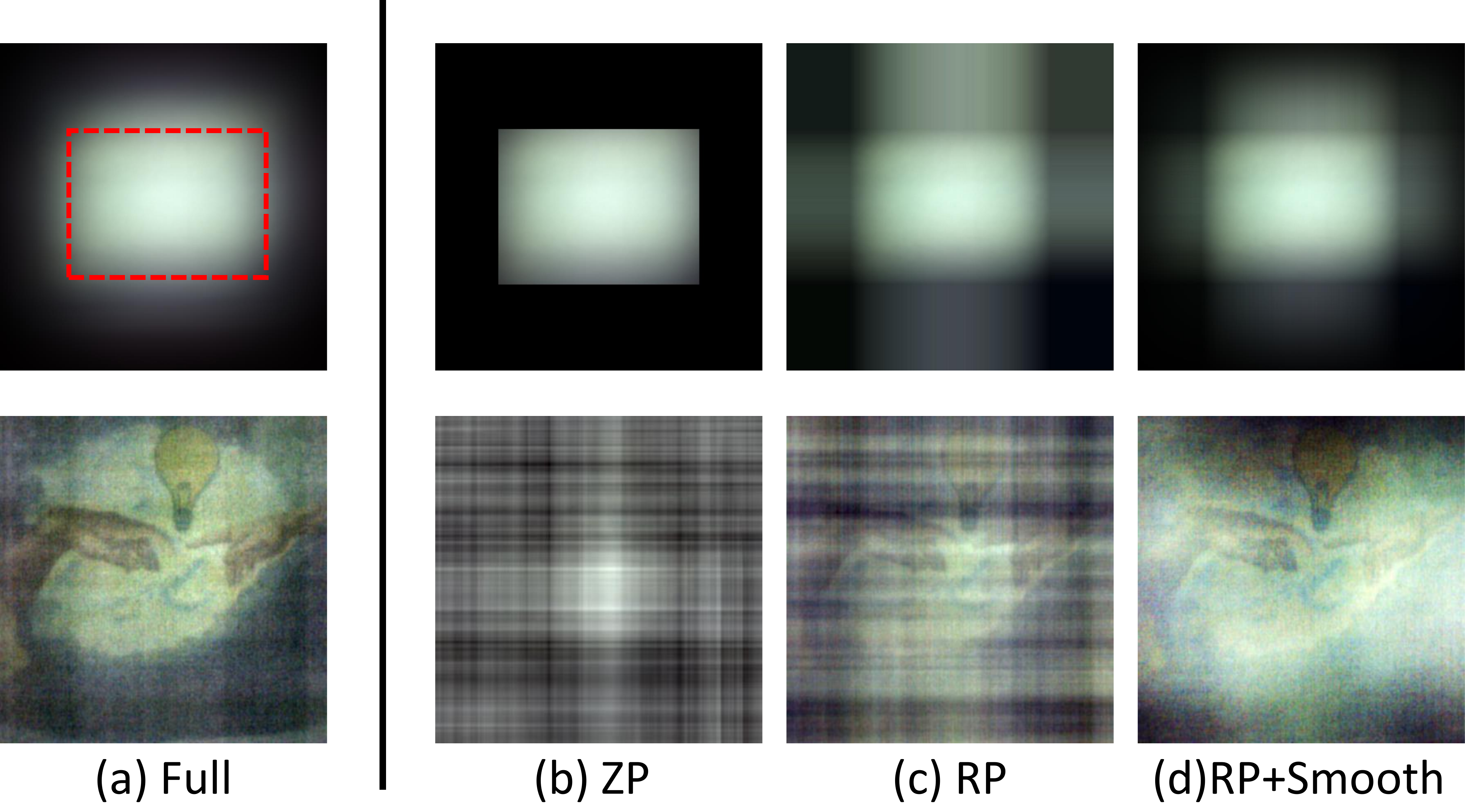}
    \begin{center}
    \caption{\textbf{Effect of padding on Wiener deconvolution for cropped measurement}. Top row shows the measurement while the bottom row shows the corresponding Wiener reconstruction. (a) Full measurement. Red box indicates the cropped out region. (b) Zero padded measurement and the corresponding reconstruction. (c) Replicate padded measurement and the corresponding reconstruction. (d) Smoothened replicate padded measurement along with the corresponding reconstruction. Line artifacts are significantly reduced in (d) which is used in this work.}
    \label{fig:padef}
    \end{center}
\end{figure}
\begin{figure*}[!ht]
    \includegraphics[scale=0.22]{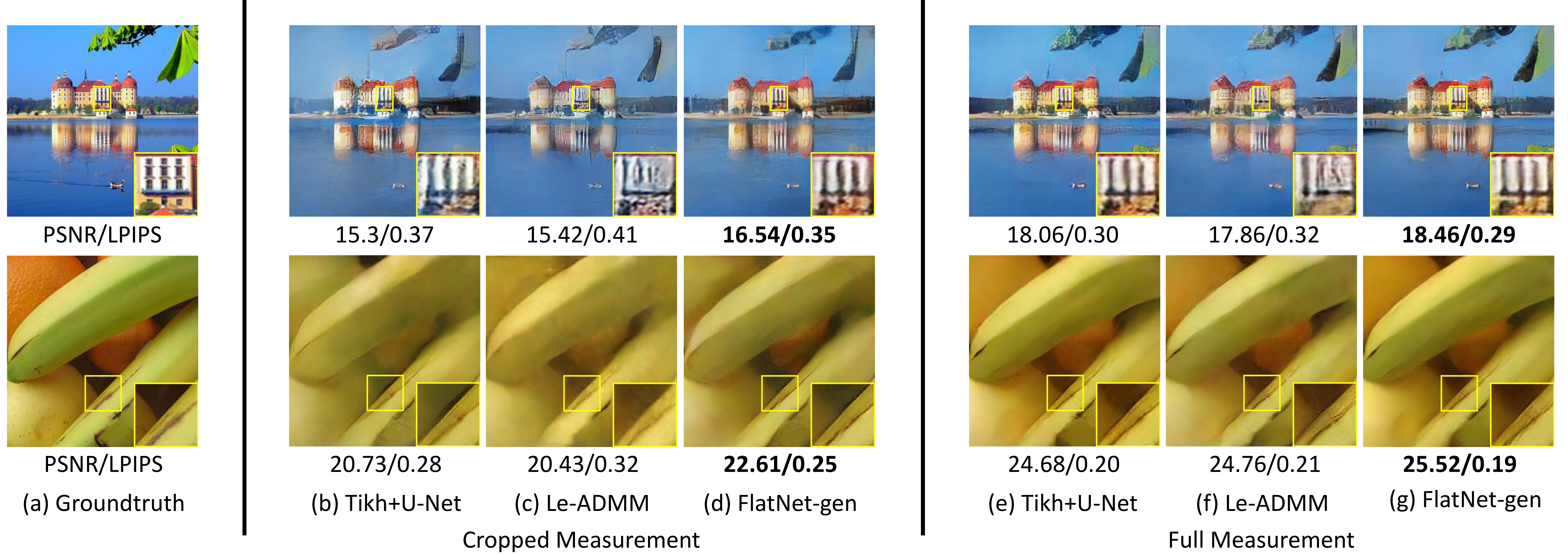}
    \begin{center}
    \caption{\textbf{Display Captured Reconstructions for cropped PhlatCam measurements}. {The difference observed in the performance of FlatNet for cropped and full measurements is small. This difference is, however, large for both Le-ADMM and Tikh+U-Net.}}
    \label{fig:rec_crop_display}
    \end{center}
\end{figure*}

\begin{figure}[!ht]
\centering
    \includegraphics[width=\columnwidth]{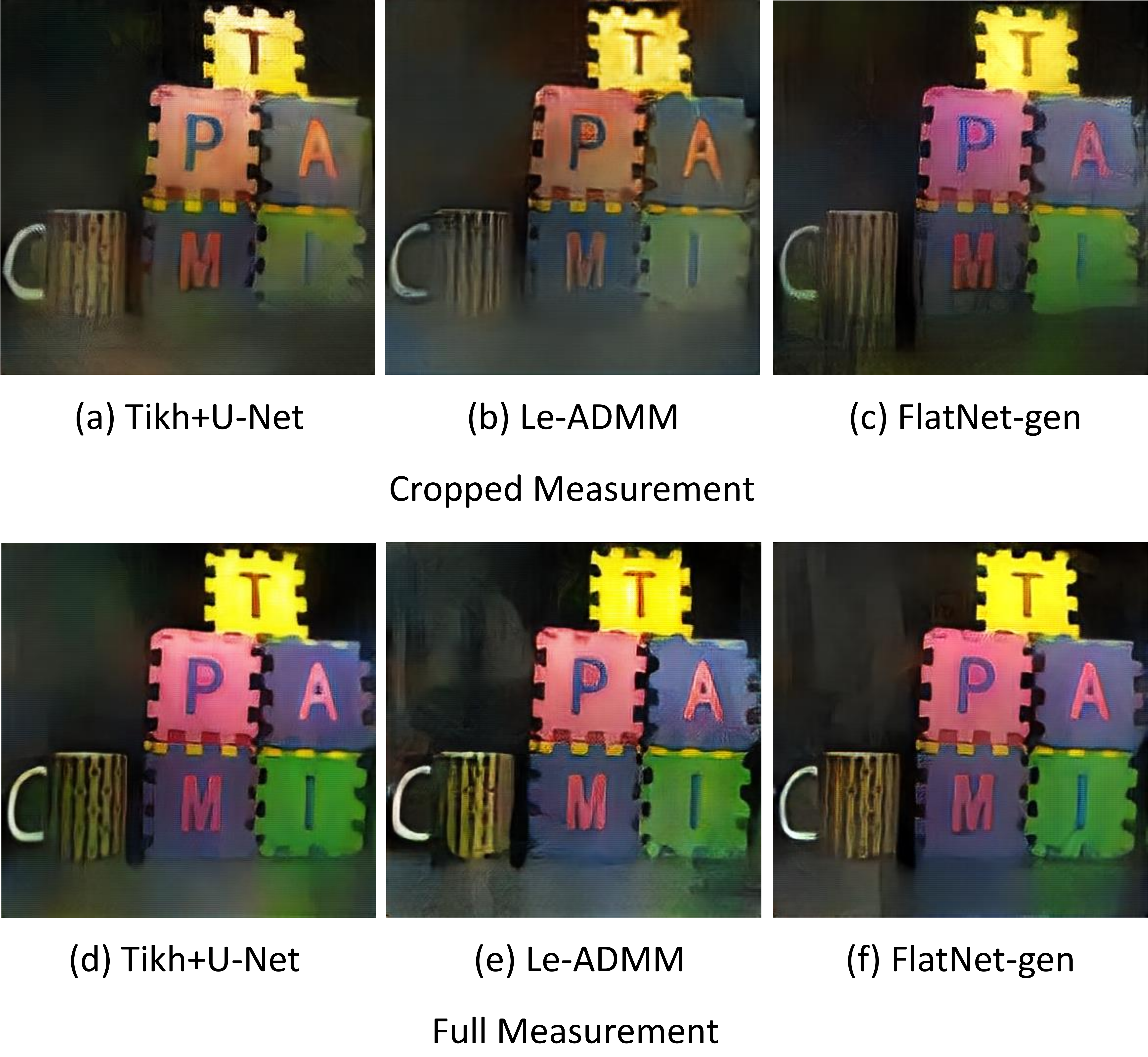}
    \begin{center}
    \caption{\textbf{Direct Captured Reconstructions for cropped PhlatCam measurements}. We can see FlatNet-gen performs reasonably well while both Le-ADMM and Tikh+U-Net breakdown. This can be observed through the colour of the letters and hazy appearance especially around the borders in Tikh+U-Net and Le-ADMM.}
    \label{fig:rec_crop_direct}
    \end{center}
\end{figure}

\begin{figure}[]
\centering
     \includegraphics[scale=0.32]{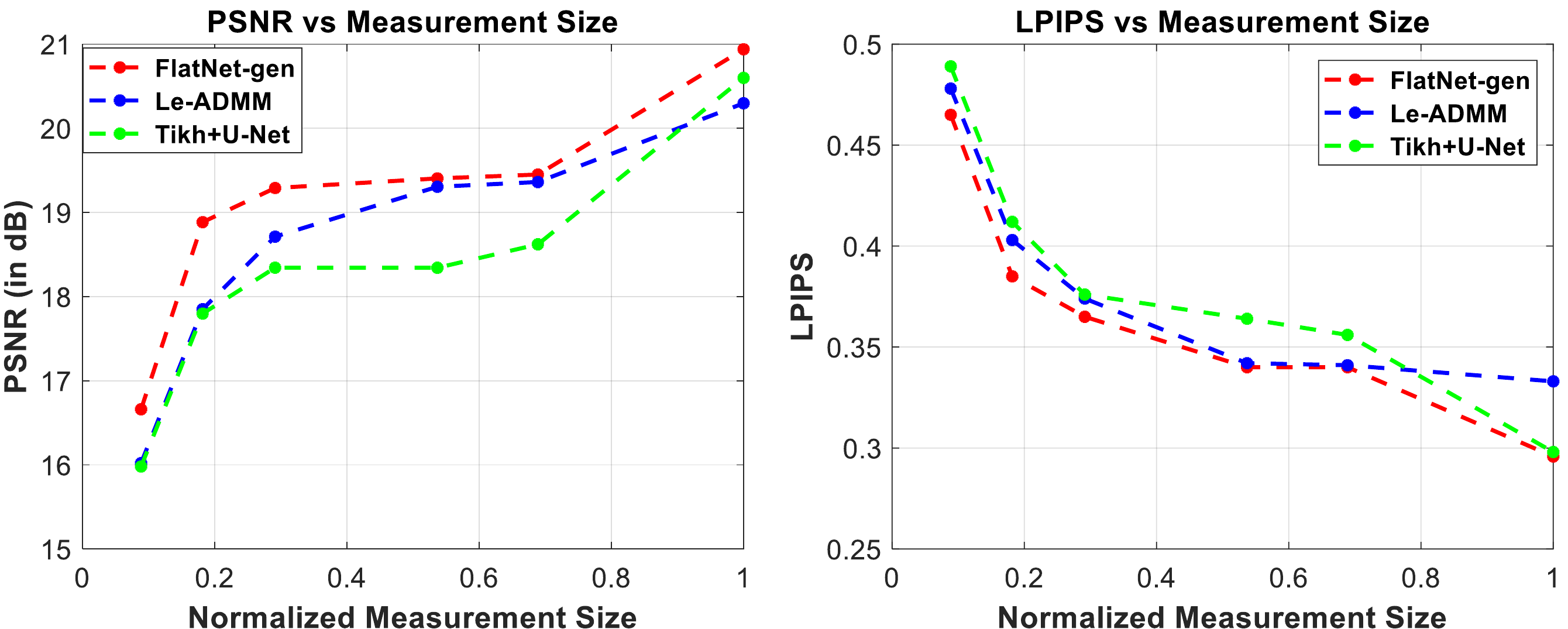}
    \begin{center}
    \caption{\textbf{Performance of learning based techniques for various amount of crops}. We plot the PSNR and LPIPS of FlatNet-gen, LeADMM and Tikh+U-Net under various measurement sizes normalized with respect to full measurement size. We can see FlatNet-gen consistently outperforms other learning based methods for all crop sizes.}
    \label{fig:crop_plot}
    \end{center}
\end{figure}

Following the observation in \cite{reeves2005fast}, we replicate pad our cropped measurements as a pre-processing step. To smooth the discontinuities due to padding, we multiply this padded measurement with a gaussian filtered box. The effectiveness of our method of padding can be observed in Figure \ref{fig:padef}. Mathematically, the trainable inversion stage changes to,
\begin{equation}\label{eq:ti_cropconv}
    {X_{\text{interm}}} = \mathcal{F}^{-1}(\mathcal{F}({W})\odot \mathcal{F}(pad({Y}))).
\end{equation}

This is a modification to Equation \ref{eq:ti_conv} to account for the cropped measurement. $pad(.)$ refers to the padding and smoothing operation described above. The same padding and smoothing procedure is also followed for Tikh+U-Net applied on the cropped measurements.
Figure \ref{fig:rec_crop_display} shows the reconstruction quality for the display captured cropped measurement compared with full measurement for Tikh+U-Net, Le-ADMM and FlatNet.
Even after padding the measurements, there are artifacts in the Wiener restored images that cannot be effectively removed using Tikh+U-Net. Le-ADMM performs slightly better than Tikh+U-Net due to its intermediate stage that approximately estimates the uncropped measurement. However, it is not as robust to crop as FlatNet-gen is. Similarly, in Figure \ref{fig:rec_crop_direct}, we show the reconstructions for direct captured cropped measurement. { It can be clearly seen that Tikh+U-Net and Le-ADMM suffer from significant color artifacts}. These artifacts are however not significant in the FlatNet-gen reconstructions. Table \ref{tab:table3} gives the comparison of average scores for each model on the display captured dataset.

It should be noted that for the model used to obtain Figures \ref{fig:rec_crop_display} and \ref{fig:rec_crop_direct} and Table \ref{tab:table3}, the PSF size ($608\times870$) exceeds the assumed sensor size ($606\times 864$). In such a case, estimation of the true PSF is a tedious process and one can use the uncalibrated FlatNet-gen-UC. From Table \ref{tab:table3}, we can see that FlatNet-gen outperforms all other learned methods. FlatNet-gen-UC has a comparable performance to FlatNet-gen, while Tikh+U-Net-UC and Le-ADMM-UC breakdown: indicating that accurate PSF calibration is required for these methods. The visual results for FlatNet-gen-UC for cropped measurements are provided in the supplementary material. 

Apart from the crop size mentioned above, we also show the performance of the learning based approaches for various different crop sizes in Figure \ref{fig:crop_plot}. Here, we normalize the size of the cropped measurements with respect to the full measurements. It can be seen that FlatNet-gen consistently outperforms Le-ADMM and Tikh+U-Net for all crop sizes.

It should also be noted that FlatNet-sep is, by design, robust to non-circulant scenarios as it involves learned inversion in the spatial domain.
\begin{figure*}[!ht]
\label{fig:dyn}
\centering
\includegraphics[scale=0.35]{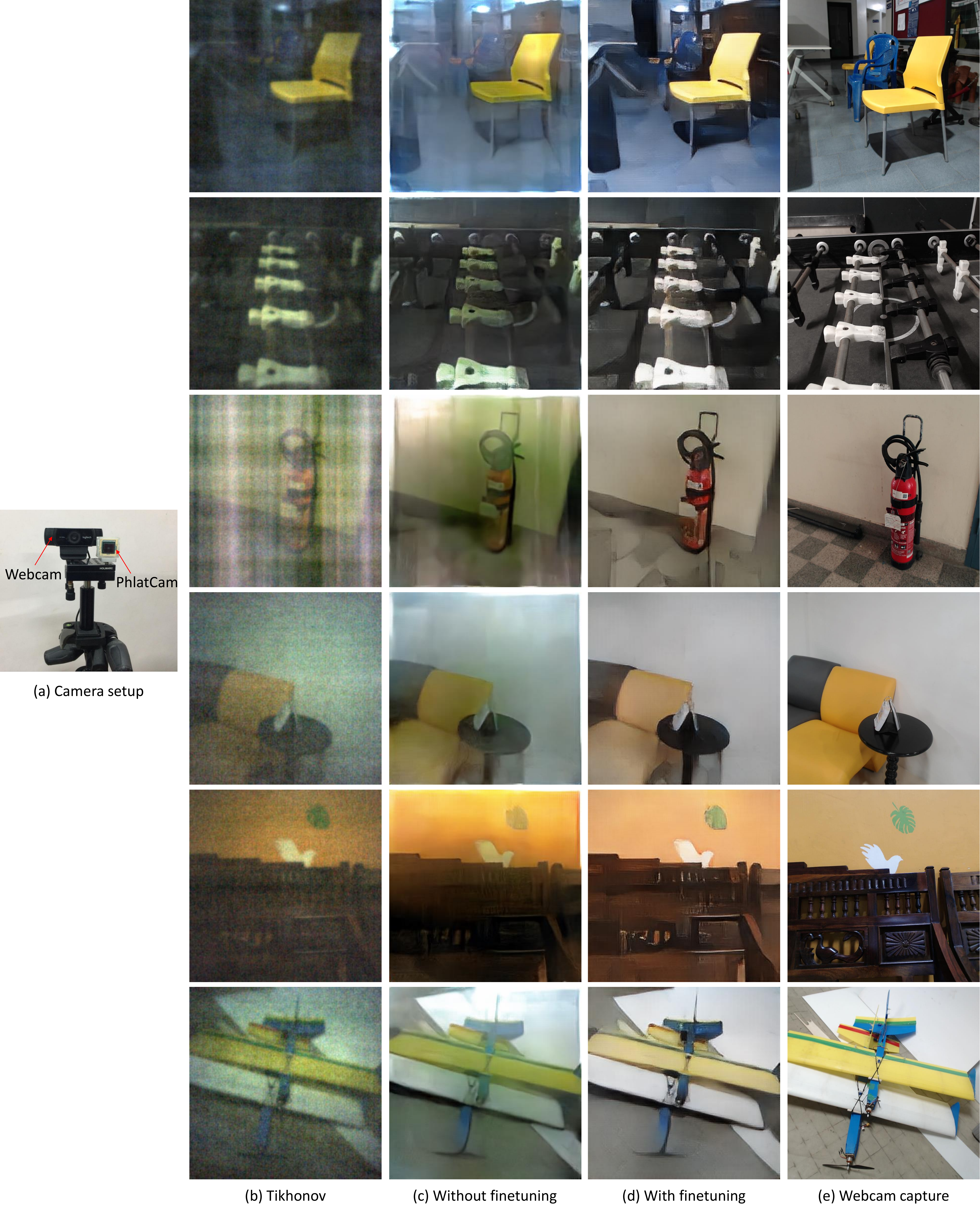}
\begin {center}
     \caption{\textbf{{Photorealistic reconstruction for unconstrained indoor scenes}}. (a) The PhlatCam-Webcam setup to capture the dataset for finetuning FlatNet-gen. (b) Tikhonov reconstruction. (c) Reconstructions from FlatNet-gen trained just on display captured data. (d) Reconstructions using FlatNet-gen finetuned on unconstrained indoor captures. (e) Webcam image for reference. Finetuning makes the reconstructions more realistic.}
    \label{fig:finetune}
    \end{center}
\end{figure*}

\subsection{Performance on unconstrained indoor scenes}\label{uncontrolled}

\begin{table}
\caption{\textbf{Average Metrics on cropped Display Captured PhlatCam measurements}. FlatNet-gen performs consistently better than other learned approaches for both real (FlatNet-gen-C) and simulated PSF case(FlatNet-gen-UC). It should be noted that FlatNet-gen-UC performs as good as Le-ADMM based on real PSF.}
\centering
\begin{tabular}{ |c||c|c|c| }
\hline
\textbf{Method}&\textbf{PSNR(in dB)} &\textbf{SSIM}& \textbf{LPIPS}\\
\hline
Tikh+U-Net-UC    & 17.53 & 0.45 & 0.438\\
\hline
Tikh+U-Net-C    & 18.34 & 0.48 & 0.376\\
\hline
Le-ADMM-UC   &  17.94 & 0.45 & 0.410\\
\hline
Le-ADMM-C   &  18.72 & 0.48 & 0.371\\
\hline
FlatNet-gen-UC    & 18.72 & 0.48 & 0.375\\
\hline
\textbf{FlatNet-gen-C}    & \textbf{19.29} &\textbf{0.50} & \textbf{0.365}\\
\hline
\end{tabular}
\label{tab:table3}
\end{table}

In the previous sections, we performed all our experiments using FlatNets trained on display captured dataset. However, real measurements captured in the wild differs from the dispay captured measurements for the following reasons: a) real world captures have significantly higher amount of noise compared to display captured measurements, b) in an unconstrained setup, bright scene points beyond the FoV described by the Chief Ray Angle (CRA) can also influence the captured measurement which is not the case with display captured measurements captured with monitors filling the whole of CRA defined FoV. To take these differences into account and make our FlatNet robust to real world scenarios, we finetune FlatNet using a real world dataset we captured called the Unconstrained Indoor Dataset. This dataset consists of unaligned webcam and PhlatCam captures collected using the setup described in Figure \ref{fig:finetune}. We collected 500 pairs of such data, keeping 475 pairs for training and 25 for testing. We finetune the entire network with a small learning rate ($10^{-12}$ for the trainable inversion stage and $10^{-6}$ for the perceptual enhancement stage). To account for misalignment between PhlatCam and webcam captures, we only use Contextual Loss\cite{mechrez2018contextual} which was previously proposed for unaligned data. Figure \ref{fig:finetune} shows some of our reconstruction results with and without finetuning along with webcam captures for reference. It can be observed that finetuning results in more photorealistic reconstructions.
In the supplementary material, we show reconstructions from cropped unconstrained indoor measurements. 


\section{Discussion and Conclusion}\label{conclusion}
In this paper, we propose an end-to-end trainable deep network called FlatNet for photorealistic scene reconstruction from lensless measurements. Despite the numerous promises that lensless imaging provides, it is somewhat restricted by the quality of the reconstructed image. In this paper, we have attempted to bridge this gap between the promise of lensless imaging and its performance. FlatNet leverages the physics of the forward model (through the trainable camera inversion) and the success of data-driven approaches to learn a photorealistic mapping from the highly multiplexed lensless captures to the estimated scene. Unlike unrolling based networks\cite{monakhova2019learned}, it has the advantage of low memory and computational requirements which are desirable criteria for stand-alone devices. We also show that by finetuning FlatNet trained on display captured measurements, using unaligned Webcam-PhlatCam indoor scenes, we can recover photorealistic images in the wild using these ultra-thin sensors. 

It should also be noted that like most GAN based approaches, FlatNet reconstructions suffer from hallucination artifacts that favor photorealism over high-fidelity. Therefore, FlatNet should be used with caution when the task at hand is critical to these hallucination artifacts (for example medical imaging). Nevertheless, in such critical systems, one can still use the trainable camera inversion of FlatNet and make modifications to the perceptual enhancement and the losses appropriately.

In future, it would be interesting to look into the co-design of mask or PSF and reconstruction algorithm for mask-based lensless cameras.


\section*{Acknowledgments}
This work was supported in part by NSF CAREER: IIS-1652633, NSF EXPEDITIONS: CCF-1730574, DARPA NESD: HR0011-17-C0026, NIH Grant: R21EY029459 and the Qualcomm Innovation Fellowship 2020 India.
\bibliographystyle{IEEEtran}



\vspace{-14mm}
\begin{IEEEbiography}[{\includegraphics[width=1.1in,height=1.2in]{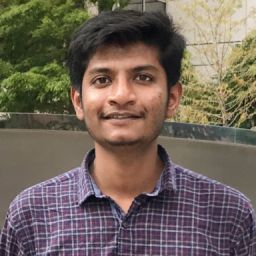}}]{Salman Siddique Khan}
is currently a Ph.D. student in the Department of Electrical Engineering, IIT Madras, India. He received the B.Tech degree in Electronics and Instrumentation Engineering from the National Institute of Technology, Rourkela, India in 2018. His research interests include computational imaging, signal processing, optics and computer vision.
\end{IEEEbiography}
\vspace{-10mm}
\begin{IEEEbiography}[{\includegraphics[width=1.1in,height=1.2in]{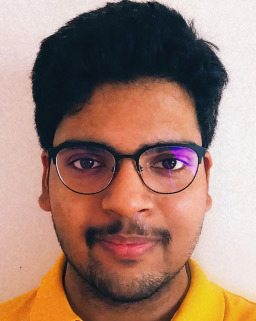}}]{Varun Sundar}
is presently an undergraduate in the Department of Electrical Engineering at IIT Madras, India. He is also an incoming PhD student at the University of Wisconsin Madison. At IIT Madras, he is associated with the Computational Imaging lab, where he worked on lensless imaging systems.
\end{IEEEbiography}
\vspace{-7mm}
\begin{IEEEbiography}[{\includegraphics[width=1.1in,height=1.2in]{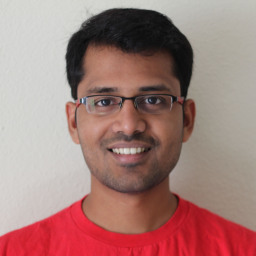}}]{Vivek Boominathan}
received the B.Tech degree in Electrical Engineering from the Indian Institute of Technology Hyderabad, Hyderabad, India, in 2012, and the M.S. and Ph.D. degree in 2016 and 2019, from the Department of Electrical and Computer Engineering, Rice University, Houston, TX, USA. He is currently a Postdoctoral Associate with Rice University, Houston, TX. His research interests lie in the areas of computer vision, signal processing, wave optics, and computational imaging.
\end{IEEEbiography}
\vspace{-7mm}
\begin{IEEEbiography}[{\includegraphics[width=1.1in,height=1.2in]{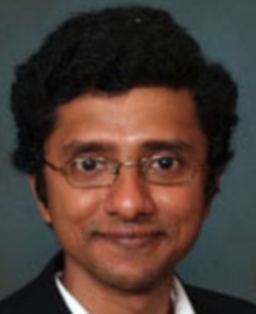}}]{Ashok Veeraraghavan}
received the bachelor’s degree in electrical engineering from the Indian Institute of Technology, Madras, Chennai, India, in 2002 and the M.S. and Ph.D. degrees from the Department of Electrical and Computer Engineering, University of Maryland, College Park, MD, USA, in 2004 and 2008, respectively. He is currently an Associate Professor of Electrical and Computer Engineering, Rice University, Houston, TX, USA. Before joining Rice University, he spent three years as a Research Scientist at Mitsubishi Electric Research Labs, Cambridge, MA, USA. His research interests are broadly in the areas of computational imaging, computer vision, machine learning, and robotics. Dr. Veeraraghavan’s thesis received the Doctoral Dissertation Award from the Department of Electrical and Computer Engineering at the University of Maryland. He is the recipient of the National Science Foundation CAREER Award in 2017. At Rice University, he directs the Computational Imaging and Vision Lab.
\end{IEEEbiography}
\vspace{-7mm}
\begin{IEEEbiography}[{\includegraphics[width=1.1in,height=1.2in]{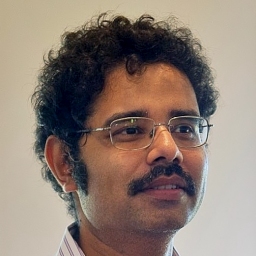}}]{Kaushik Mitra}
received the Ph.D. degree from the Department of Electrical and Computer Engineering, University of Maryland, College Park, MD, USA. He is currently an Assistant Professor with the Department of Electrical Engineering, Indian Institute of Technology Madras, Chennai, India. Before joining IIT Madras, he was a Postdoctoral Research Associate with the Department of Electrical and Computer Engineering, Rice University, Houston, TX, USA. His research interests include computational imaging, computer vision, and machine learning. His contributions to computational imaging include proposing a theoretical framework for analysis and design of novel computational imaging systems, development of novel imaging systems, such as hybrid light field camera and assorted camera array, and using machine learning techniques, such as dictionary learning and deep learning for improving the performance of computational imaging systems.
\end{IEEEbiography}


\end{document}



\IEEEtitleabstractindextext{
\begin{abstract}
In this supplementary material, we provide some additional details. We provide details about the display captured setup, the qualitative performance of FlatNet-gen-UC on both cropped and full measurements, the variation of performance of the deep networks with respect to the number of parameters, additional detail on the trainable inversion stage, the performance of FlatNet-gen finetuned on unconstrained cropped indoor captures and the performance of both FlatNet-sep and FlatNet-gen on scenes containing bright objects.
\end{abstract}
\begin{IEEEkeywords} 
lensless imaging, image reconstruction 
\end{IEEEkeywords}
}

\maketitle

\section{Display Capture Setup}
\begin{figure}
    \centering
    \includegraphics[scale=0.1]{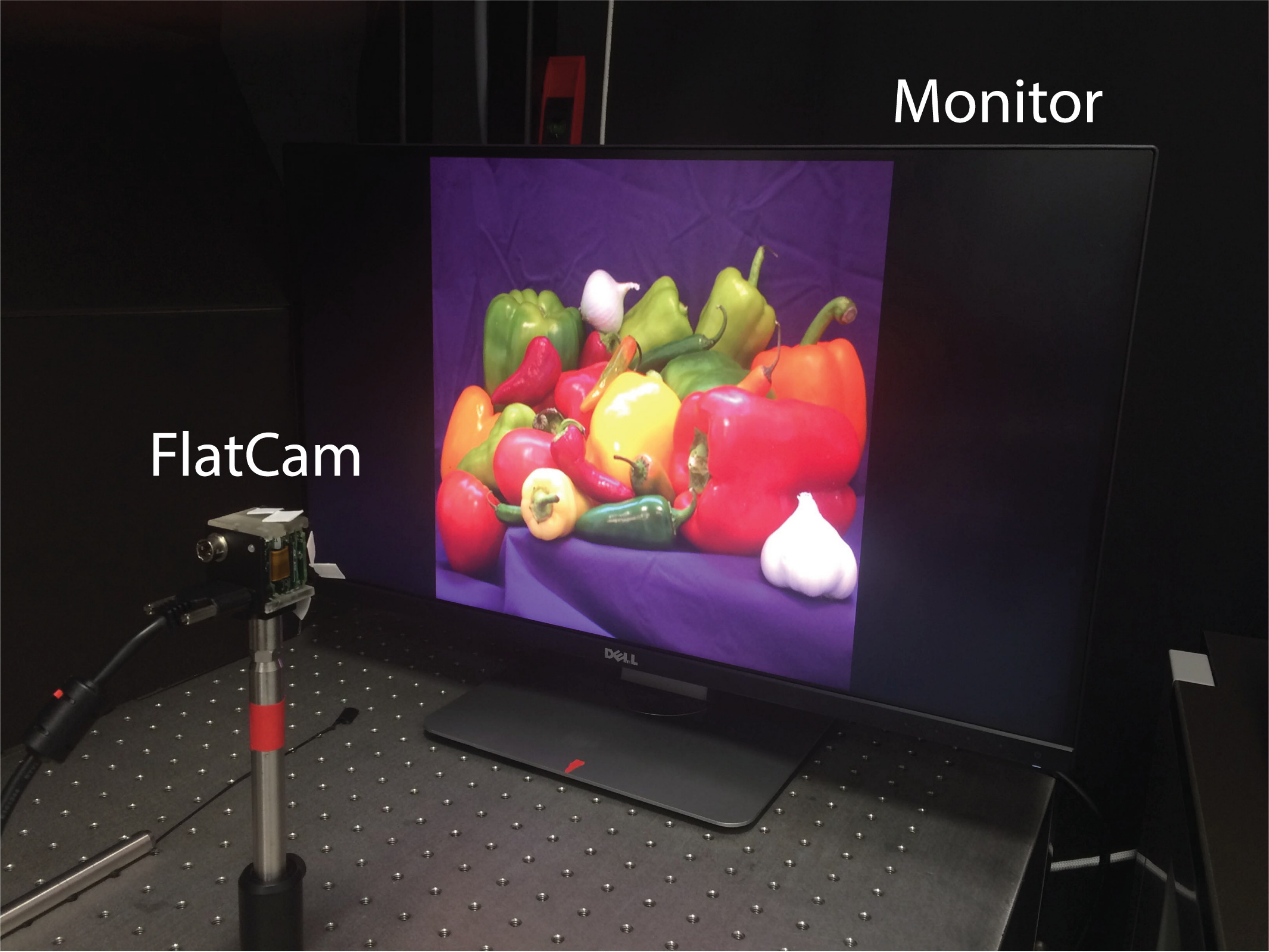}
    \caption{\textbf{The display capture setup for FlatCam.} A similar setup was used for PhatCam.}
    \label{fig:setup}
\end{figure}

To capture a display-captured image using FlatCam\cite{asif2017flatcam} and PhlatCam\cite{boominathan2020phlatcam}, the image is resized so as to occupy the biggest central square on a 24-inch monitor using bicubic interpolation. The monitor was placed at appropriate distance so that the image occupied the field of view of the cameras. For FlatCam, this was around 1 foot, while for PhlatCam, this was around 16 inches. This setup is fixed for all image captures such that the alignment of the monitor pixels to the camera pixels is uniform throughout both training and test. The white balance setting for FlatCam is fixed to be the white balance setting obtained in the FlatCam’s (i.e. PointGrey Flea3) automatic white balance mode when an all-white image is displayed on the monitor. The exposure time is set to PointGrey’s automatic mode, and the camera’s gain is set to 0dB. For PhlatCam prototype using a Basler ace camera, the white balance setting was estimated once before the capture began by capturing a demo picture. The exposure was set at 10000 microseconds. Figure \ref{fig:setup} shows the setup for FlatCam capture. The setup for PhlatCam is similar.


\section{Qualitative Comparison for Uncalibrated PSF Case}
\begin{figure*}[!ht]
    \includegraphics[scale=0.22]{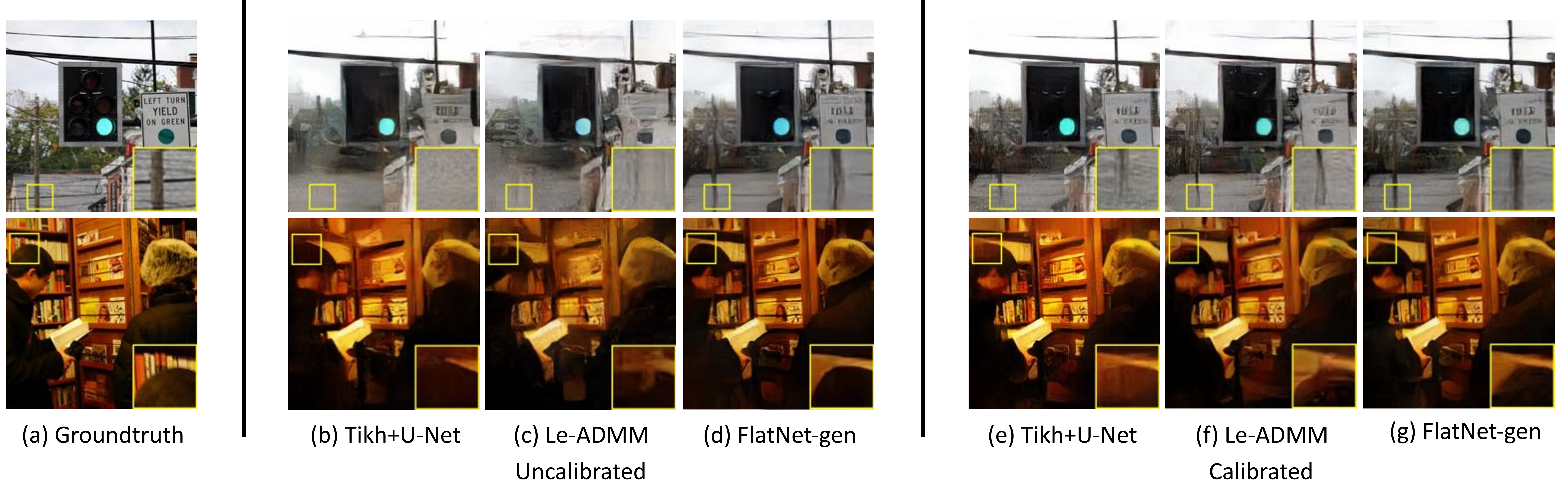}
    \begin{center}
    \caption{\textbf{Comparison between uncalibrated and calibrated learning based approaches for full PhlatCam measurement}. Tikh+U-Net and Le-ADMM rely on accurate estimation of PSF while FlatNet-gen relies on PSF only for initialization and rather learns the inverse of the PhlatCam forward model. FlatNet-gen higher quality reconstructions with finer details for both calibrated and uncalibrated case. This is not the case for Le-ADMM or Tikh+U-Net.}
    \label{fig:uc_full_disp}
    \end{center}
\end{figure*}

\begin{figure*}[!ht]
    \includegraphics[scale=0.22]{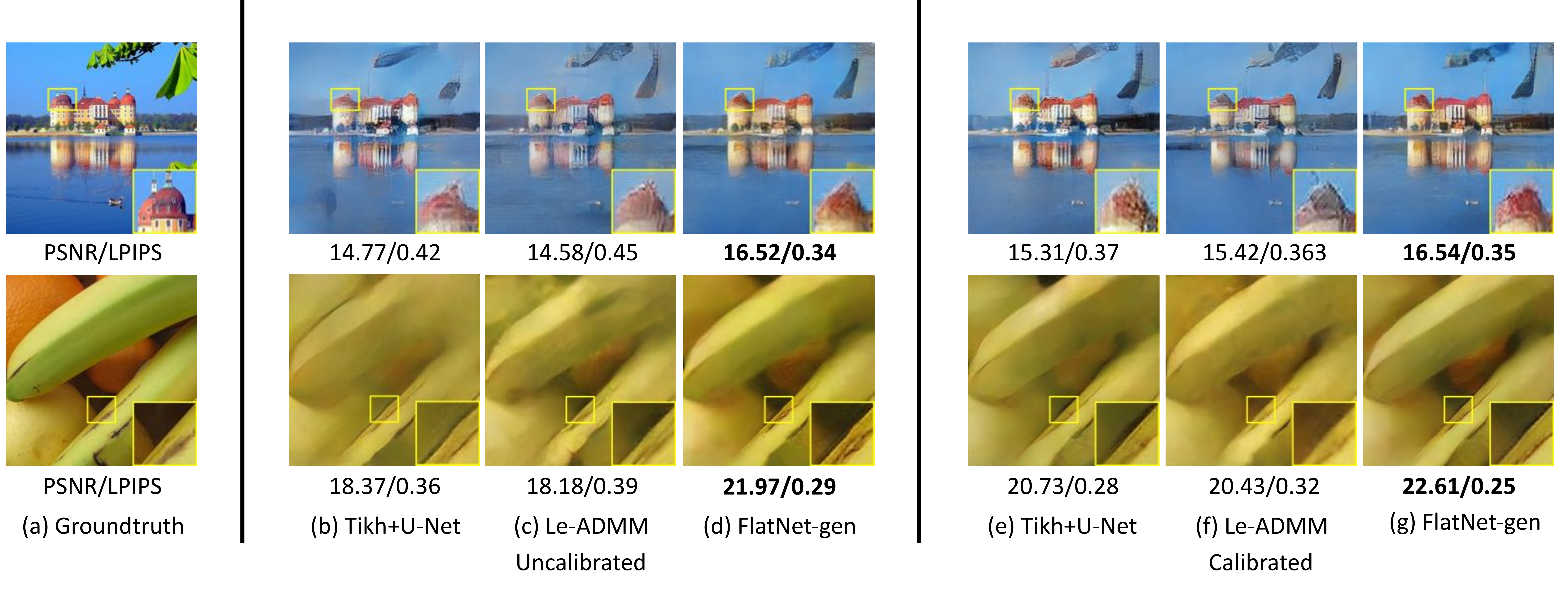}
    \begin{center}
    \caption{\textbf{Comparison between uncalibrated and calibrated learning based approaches for cropped PhlatCam measurement}. FlatNet-gen provides higher quality reconstruction for both calibrated and uncalibrated case even when the measurement is extensively cropped. This indicates that FlatNet-gen can be used for small sensor setup without accurately estimating the PSF.}
    \label{fig:uc_crop_disp}
    \end{center}
\end{figure*}
In Section 4.3.2 and 4.4.1 of the main paper, we provided the quantitative comparison for FlatNet-gen with Le-ADMM and Tikh+U-Net. In this section, we provide the visual results for the uncalibrated versions of the same. In particular, we use PSF simulated using the method described in Section 3.1.2 and use this PSF for learning Le-ADMM, Tikh+U-Net and FlatNet-gen. We provide the comparison for both full measurement in Figure \ref{fig:uc_full_disp} and cropped measurement in Figure \ref{fig:uc_crop_disp}. We can see clearly that the performance of FlatNet-gen-UC is very close to its calibrated counterpart i.e. FlatNet-gen-C. However, this is not the case with Le-ADMM and Tikh+U-Net. 


\section{Effect of parameters on performance of FlatNet-gen}
In this section, we investigate how FlatNet-gen compares against Le-ADMM and Tikh+U-Net in terms of performance for different parameter count. In particular, we train FlatNet-gen, Tikh+U-Net and Le-ADMM for different variants of U-Net, keeping the number of learnable parameters constant in the trainable inversion stage for FlatNet-gen and unrolled ADMM block for Le-ADMM. U-Net-N refers to the variant of U-Net for which the number of filters in a convolutional block increases from N to 8N and reduces back to N. We perform this experiment for N = 32, 64 and 128. Table \ref{tab:param} provides the variation of the average PSNR and LPIPS for Tikh+U-Net, Le-ADMM and FlatNet-gen against the total number of learnable parameters. It is clear that FlatNet-gen outperforms both Tikh+U-Net and Le-ADMM for different parameter counts at the cost of slight increase in the relative number of learnable parameters. In the main text, we report the best model for each approach i.e. with U-Net-128.
\begin{table}[ht]
\vspace{1.5em}
\centering
\caption{\textbf{Variation of performance against the total number of learnable parameters}. FlatNet-gen outperforms both Le-ADMM and Tikh+U-Net under all parameter counts.} 
\begin{tabular}{|c|ccc|}
 \hline
 \textbf{Methods} & \textbf{PSNR (in dB)} & \textbf{LPIPS} & \textbf{Learnable Parameters}\\ \hline
\textbf{Tikh+U-Net} &&&\\ \hline
U-Net-32 & 18.74 & 0.384 & 2.4M\\
U-Net-64 & 19.83 & 0.341 & 12.9M\\
U-Net-128 & 20.60 & 0.298 & 51.5M\\
\hline
\textbf{Le-ADMM} &&&\\ \hline
U-Net-32 & 15.72 & 0.448 & 2.4M\\ 
U-Net-64 & 17.20 & 0.407 & 12.9M\\
U-Net-128 & 20.29 & 0.333 & 51.5M\\
 \hline
 \textbf{FlatNet-gen} &&&\\ \hline
U-Net-32 & 18.83 & 0.379 & 4.2M\\ 
U-Net-64 & 19.92 & 0.336 & 14.7M\\
\textbf{U-Net-128} & \textbf{20.94} & \textbf{0.296} & \textbf{53.3M}\\
\hline
\end{tabular}
\vspace*{3mm}
\label{tab:param}
\end{table}


\section{Details of trainable camera inversion} In this section, we provide some additional details regarding the trainable camera inversion stage.
\subsection{Initial weights in FlatNet-sep}
\begin{figure}[ht]
    \centering
    \includegraphics[scale=0.19]{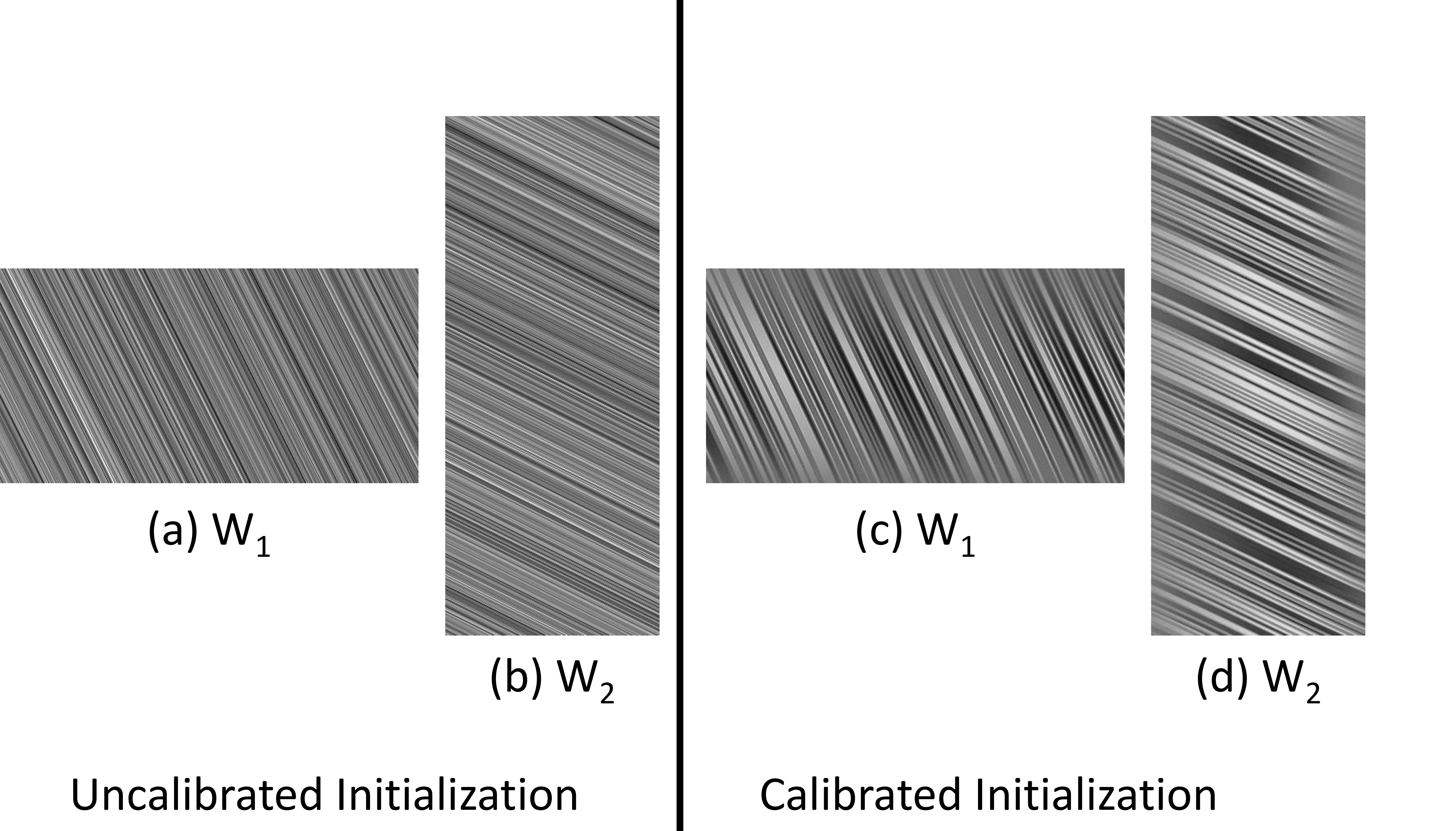}
    \caption{\textbf{Initialized trainable inversion weights for FlatNet-sep}. (a) Initialized $W_1$ for uncalibrated case. (b) Initialized $W_2$ for uncalibrated case. (c) Initialized $W_1$ for calibrated case. (d) Initialized $W_2$ for calibrated case.}    \label{fig:initWsep}
\end{figure}

The dimensions of $W_1$ and $W_2$ are $256\times 500$ and $620\times 256$, given that the measurement dimensions are $500\times 620\times 4$ and the reconstruction dimensions are $256\times 256$. For calibrated initialization of FlatNet-sep, we use $\Phi_L^T$ to initialize $W_1$ and $\Phi_R$ to initialize $W_2$. Similarly, for the uncalibrated initialization, we first generate random toeplitz matrices of slope that matches that of $\Phi_L$ and $\Phi_R$. Once these matrices are generated, they are used for initialization in a way similar to the calibrated case i.e. the transpose of the random toeplitz matrix corresponding to the $\Phi_L$ is used to initialize $W_1$ and the random toeplitz matrix corresponding to $\Phi_R$ is used to initialize $W_2$. Figure \ref{fig:initWsep} presents a visual representation of how the initialized weights $W_1$ and $W_2$ look for both calibrated and uncalibrated case.
\subsection{Generation of random toeplitz matrices for FlatNet-sep}
\begin{figure*}[!ht]
    \centering
    \includegraphics[scale=0.15]{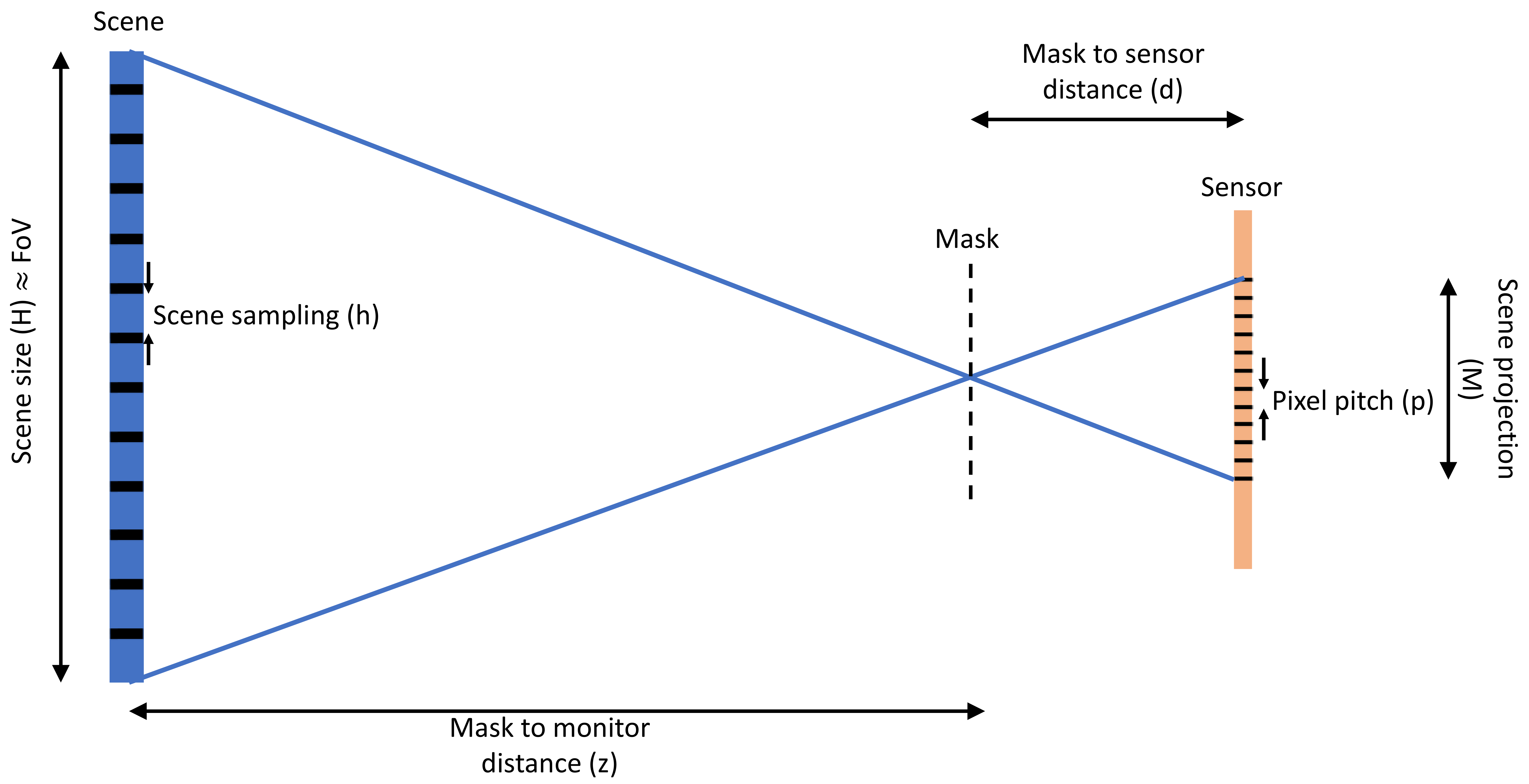}
    \caption{\textbf{Geometry considered for the generation of random toeplitz matrices}. 1-D representation of the geometry considered for the generation of matrices for uncalibrated FlatNet-sep.}    \label{fig:rand_geom}
\end{figure*}
For this subsection, please refer to Figure \ref{fig:rand_geom} which provides a 1-D version of the geometry we are considering. Let us assume that a scene of dimension $H\times W$ fills up the entire FoV of the camera and the the scene is discretized into $h\times w$ dimensional pixels. The corresponding scene maps to a region of dimension $M\times N$ in the sensor and the sensor has a pixel pitch of $p$. The slope of the calibration matrix $\Phi_L$ is then defined as follows, 
\begin{equation}\label{eq:slopeti}
    m_{L} = \frac{H/h}{M/p}
\end{equation}
This slope measures the ratio of the number of pixels (row) in the scene to the number of pixels (rows) in its projection at the sensor or in other words, how many row pixels in the scene correspond to a row pixel at the sensor.

\begin{figure}[!ht]
    \centering
    \includegraphics[scale=0.19]{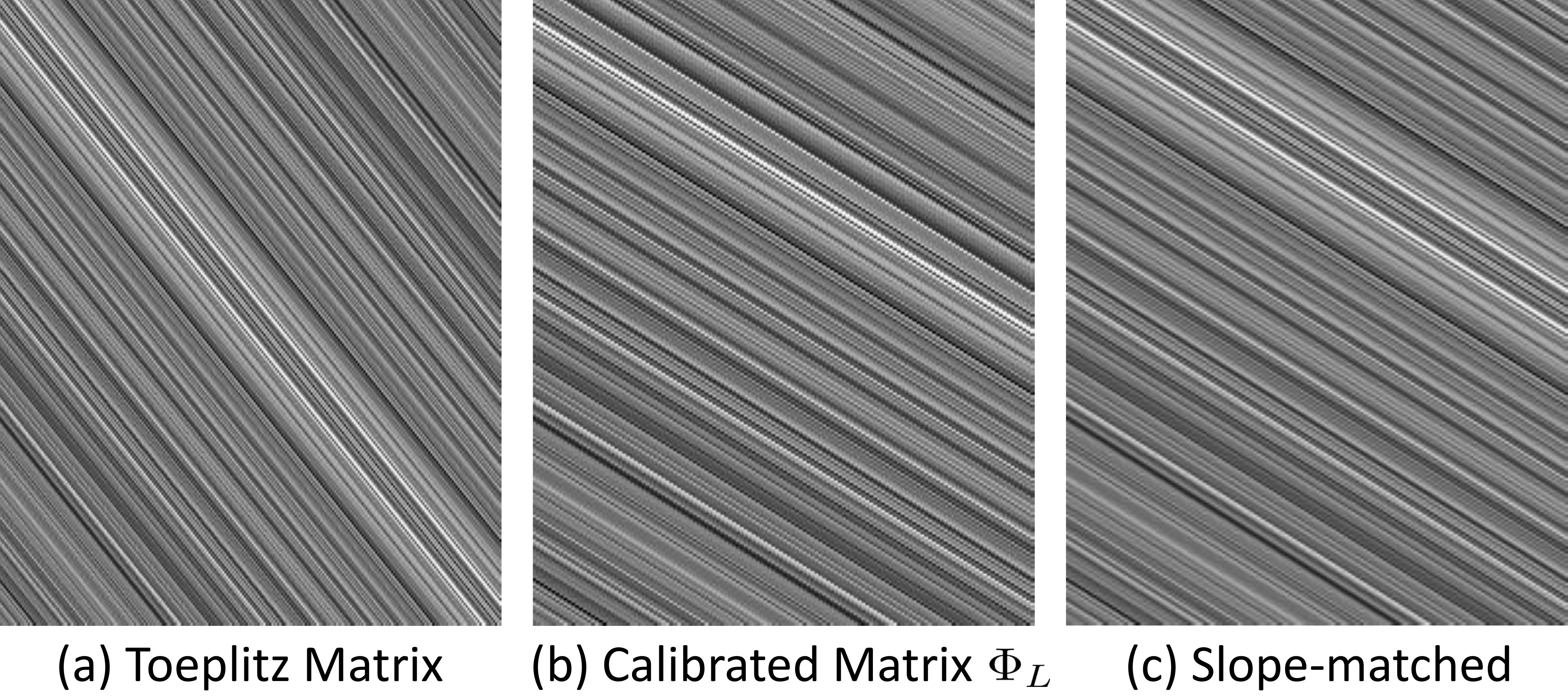}
    \caption{\textbf{Generation of slope-matched random toeplitz matrices}. (a) Shows a conventional random toeplitz matrix. (b) Shows the calibrated matrix $\Phi_L$. (c) Shows the slope-matched random toeplitz matrix. The slope in (c) matches to that in (b).}    \label{fig:slope}
\end{figure}
If we assume our monitor for calibration or data capture is at $z$ distance from the camera and the mask to sensor distance is $d$, then,
\begin{equation}\label{eq:M_ti}
    M = Hd/z
\end{equation}
Plugging \ref{eq:M_ti} into \ref{eq:slopeti}, and assuming a scene of dimension $P\times Q$ pixels (i.e. $(H/h)\times (W/w)$), the slope for $\Phi_{L}$ becomes
\begin{equation}\label{slopl}
    m_{L}  = \frac{P}{Hd/(pz)}
\end{equation}
Similarly, the slope for $\Phi_R$ can be shown to be,
\begin{equation}\label{slopr}
    m_{R} = \frac{Q}{Wd/(pz)}
\end{equation}
For the FlatCam prototype we use in the experiments, $p = 10.6\mu m$ (this pixel pitch is for each channel and is therefore twice the actual pixel pitch of the sensor) and $d = 1.5mm$. We placed the monitor at a distance $z = 31.75cm$ and projected on the screen, a scene of dimension $H \times W = 29cm\times 29cm$. If we assume our scene reconstruction to be of size $P\times Q = 256\times 256$ pixels, then $m_{L} = m_{R} \approx 2$. 

To generate toeplitz matrix of shape $S\times P$ where $P<S$ and with a slope that matches that of $\Phi_{L}$, we first generate a random vector of length $S$ and form a circulant matrix of dimension $S\times S$ corresponding to it. Then, using bilinear/nearest-neighbor interpolation, we resize this circulant matrix to $S\times m_{L}S$. We then arbitrarily crop a submatrix of size $S \times P$ from the resized matrix to match the dimension of $\Phi_L$. Similar process is followed for generating a toeplitz matrix that matches the dimension and slope $\Phi_R$ as well. Figure \ref{fig:slope} shows an example toeplitz matrix along with the calibrated $\Phi_L$ matrix and the generated slope-matched random toeplitz matrix after estimating the slope using \ref{slopl}.
\subsection{Evolution of the parameters}
\begin{figure*}
    \centering
    \includegraphics[width=\textwidth]{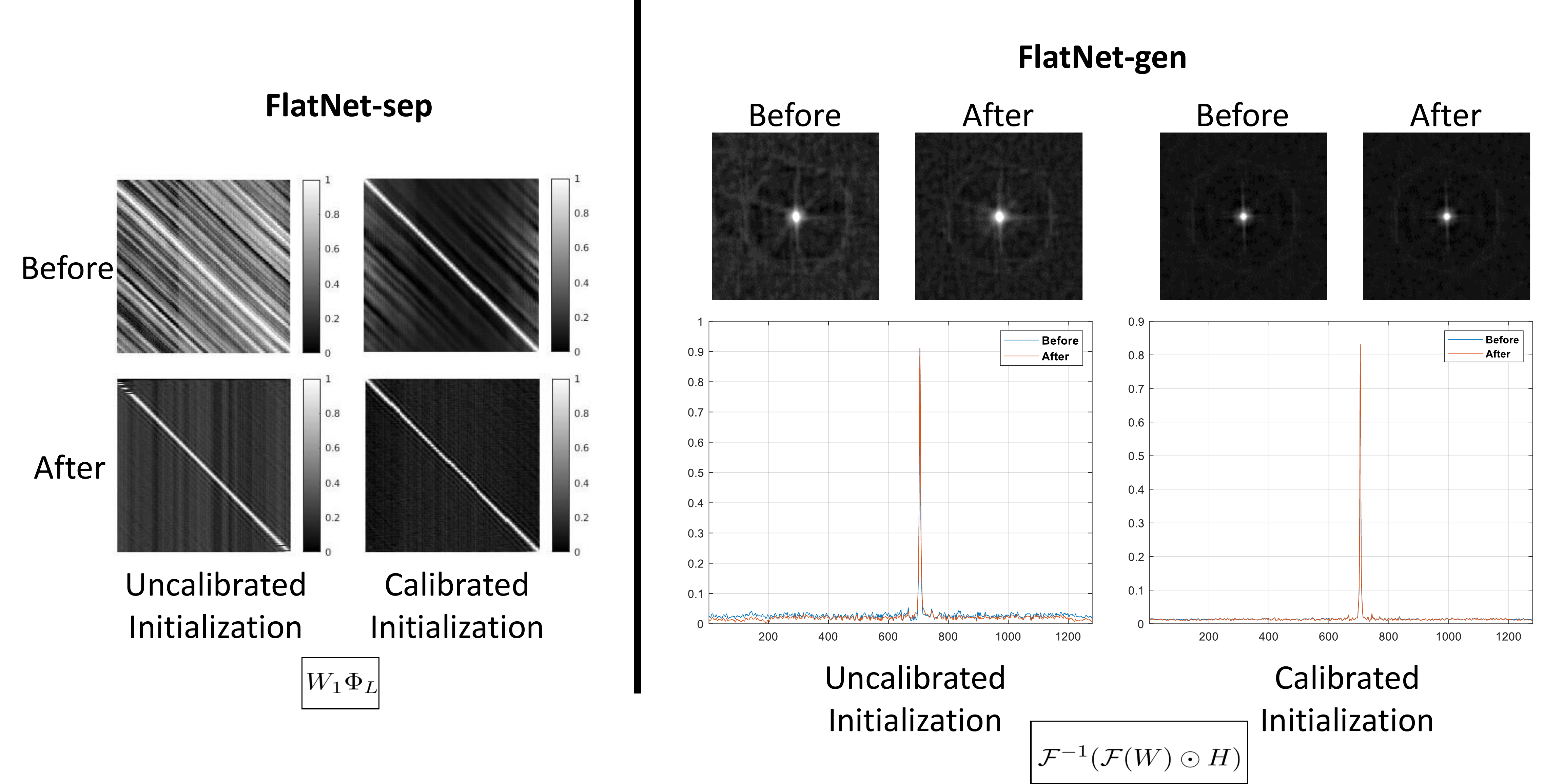}
    \caption{\textbf{Evolution of trainable camera inversion stage}. Left: $W_{1}\Phi_{L}$ is shown as an image for both uncalibrated and calibrated scenario for the inversion layer of FlatNet-sep. Eventually, the product becomes an identity matrix, indicating that the learning has led to an inversion of the forward model for FlatNet-sep. Right: $\mathcal{F}^{-1}(\mathcal{F}(W)\odot H)$ is shown for the inversion stage of both uncalibrated and uncalibrated scenario for the inversion layer of FlatNet-gen. Here $H$ is the Fourier transform of the PSF. Learning helps $W$ in inverting the PSF resulting in the impulse shown in the top figures. The bottom row shows a horizontal slice from the impulse image. The effect of learning is more prominent in the case for uncalibrated FlatNet-gen compared to the calibrated counter part due to the superior nature of the mask and the resulting Wiener filter for the calibrated case.}    \label{fig:tilayer}
\end{figure*}

\begin{figure}
    \centering
    \includegraphics[width=\columnwidth]{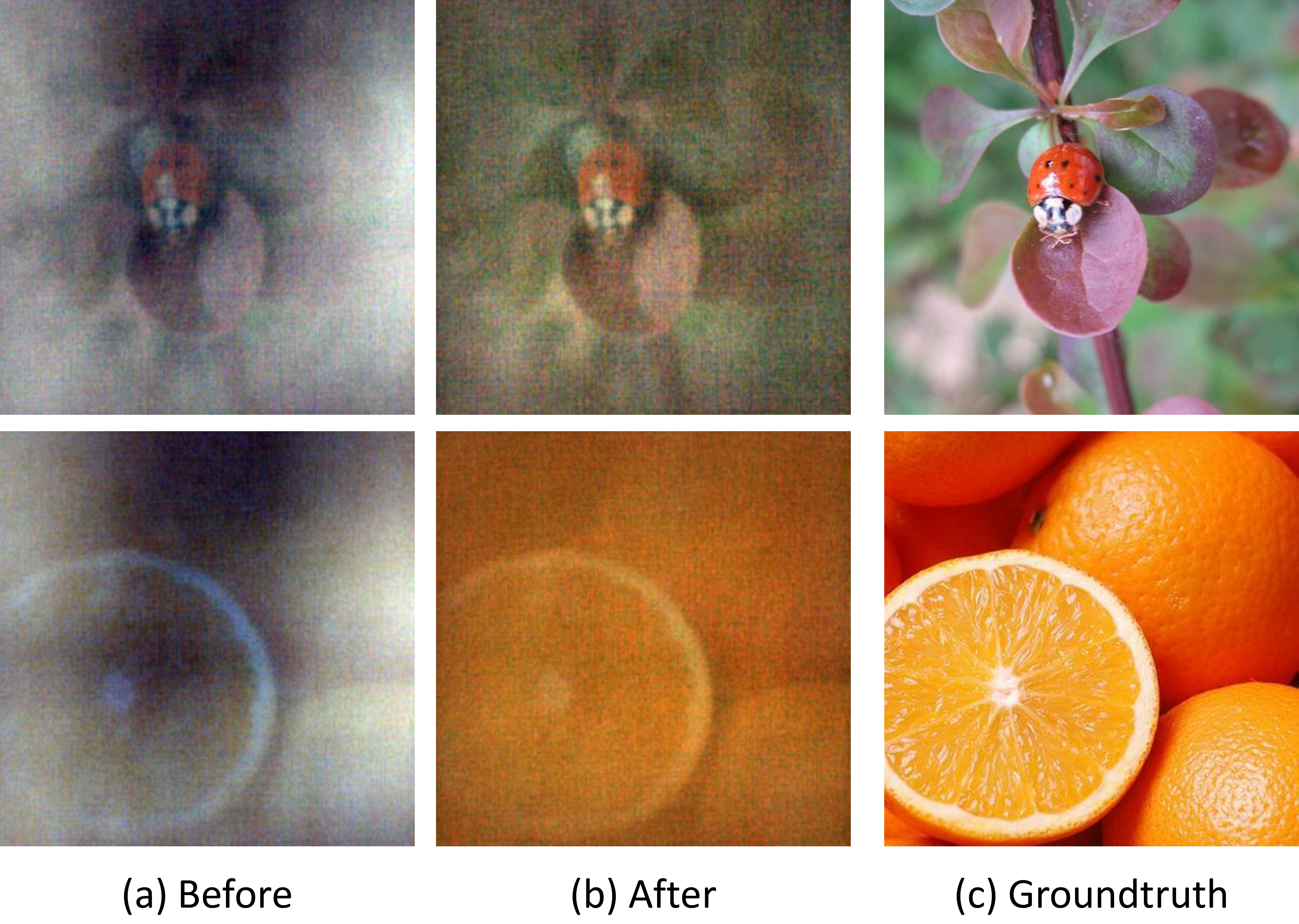}
    \caption{\textbf{Evolution of trainable inversion output of FlatNet-gen for cropped measurement}.The effect of learning of trainable inversion is more prominent for cropped measurement as can be seen here. In (a) we show the trainable inversion output at the beginning of training and in (b) we show the trainable inversion output at the end of training. It can be observed that learning has removed a majority of the artifacts. (c) Groundtruth is also shown for reference.}    \label{fig:crop_ti}
\end{figure}

Figure \ref{fig:tilayer} shows the evolution of trainable inversion parameters for both FlatNet-sep and FlatNet-gen. Specifically, we plot the product $W_{1}\Phi_{L}$ for FlatNet-sep and the convolution output $\mathcal{F}^{-1}(\mathcal{F}(W)\odot H)$ for FlatNet-gen. Here, $H$ is the Fourier transform of the PSF. For, FlatNet-sep the product is an identity matrix while the convolution output for FlatNet-gen is close to an impulse, indicating that the trainable camera inversion has learned to invert the forward process. The effect of learning is more prominent in FlatNet-sep compared to FlatNet-gen for two reasons: (a) the weights $W_1$ and $W_2$ were initialized with adjoint of $\Phi_L$ and $\Phi_R$ as compared to the pseudo-inverse of the PSF in case of FlatNet-gen, (b) owing to the superior mask properties of PhlatCam, the pseudo-inverse of the PSF is of high quality already. Similarly, the effect of learning is more prominent in case of uncalibrated initialization for FlatNet-gen compared to the calibrated counterpart. This is again due to the fact that pseudo-inverse of the calibrated PSF accurately inverts the forward model while the pseudo-inverse of the simulated PSF is unable to capture some of the non-idealities of the capturing process. As a result, it gets refined through learning to accurately invert the forward model. The prominence of the inversion stage learning in FlatNet-gen, however, is evident in the case of cropped measurements (main text section 4.4.2), as shown in Figure \ref{fig:crop_ti}. It can be seen that learning gets rid of majority of the artifact making it easier for the perceptual enhancement to extract meaningful features that help with higher quality final reconstruction.

\subsection{Additional intermediate reconstructions}
\begin{figure*}
    \centering
    \includegraphics[scale=0.22]{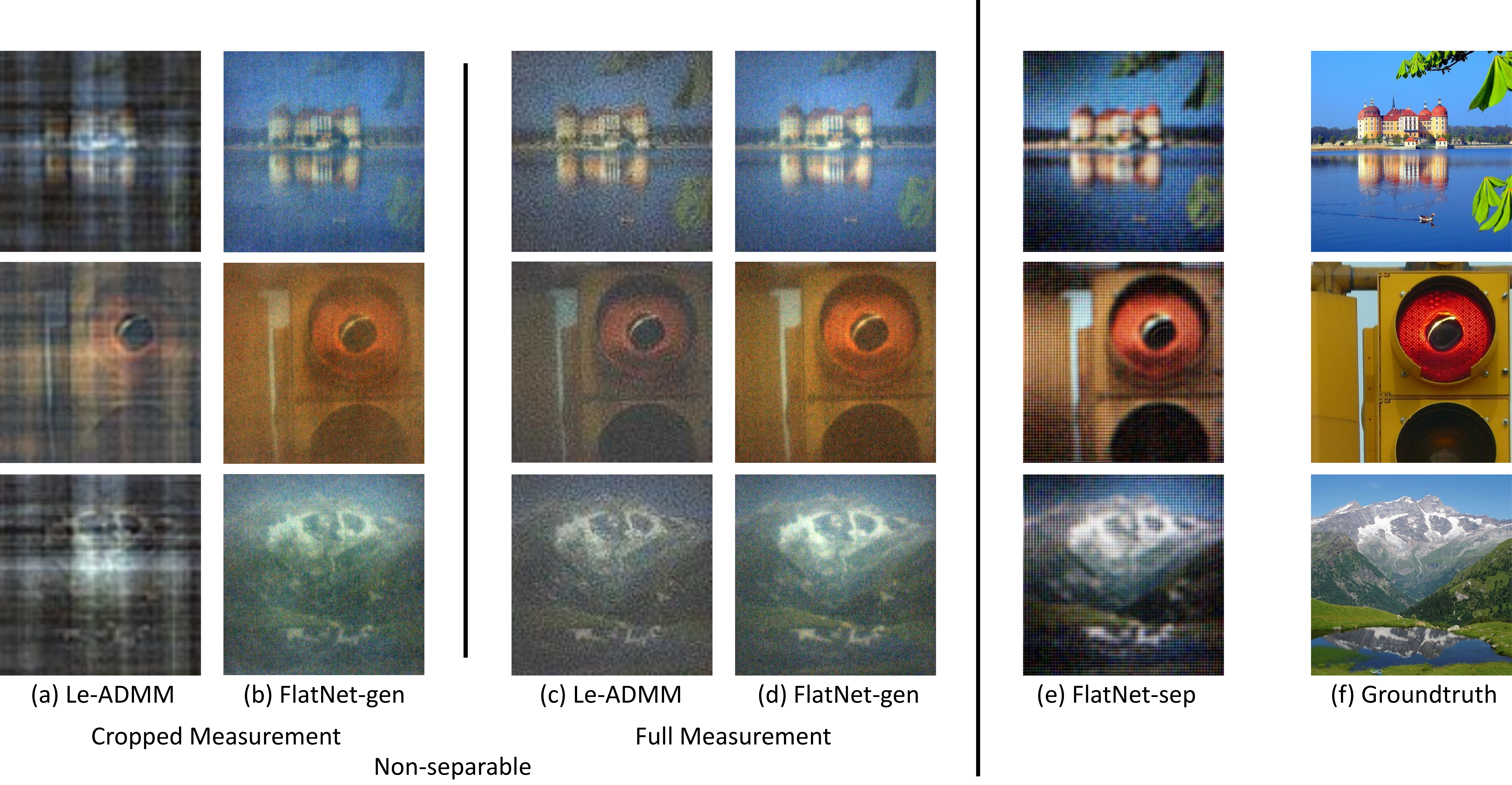}
    \caption{\textbf{Intermediate outputs before perceptual enhancement block}. (a) Intermediate output of Le-ADMM for cropped measurement. (b) Trainable inversion output of FlatNet-gen for cropped measurement. (c) Intermediate output of Le-ADMM for full measurement. (d) Trainable inversion output of FlatNet-gen for full measurement. (e) Trainable inversion out of FlatNet-sep. (f) Groundtruth for reference.}    \label{fig:ti_interm}
\end{figure*}

In this subsection, we present more intermediate results for Le-ADMM, FlatNet-gen and FlatNet-sep. In Figure \ref{fig:ti_interm}, we show the intermediate outputs for three scenes by Le-ADMM, FlatNet-gen and FlatNet-sep. The intermediate output for Le-ADMM corresponds to the output of the unrolled ADMM block while that for FlatNet corresponds to the output of the trainable inversion block. For the non-separable models (Le-ADMM and FlatNet-gen), we show the intermediates for both cropped and full measurement. We can clearly see that learning has significant impact on the intermediates especially for FlatNet-gen trained on cropped measurements.

\section{Reconstruction of unconstrained indoor scenes for small sensor}
\begin{figure*}[!ht]
\centering
    \includegraphics[scale=0.3]{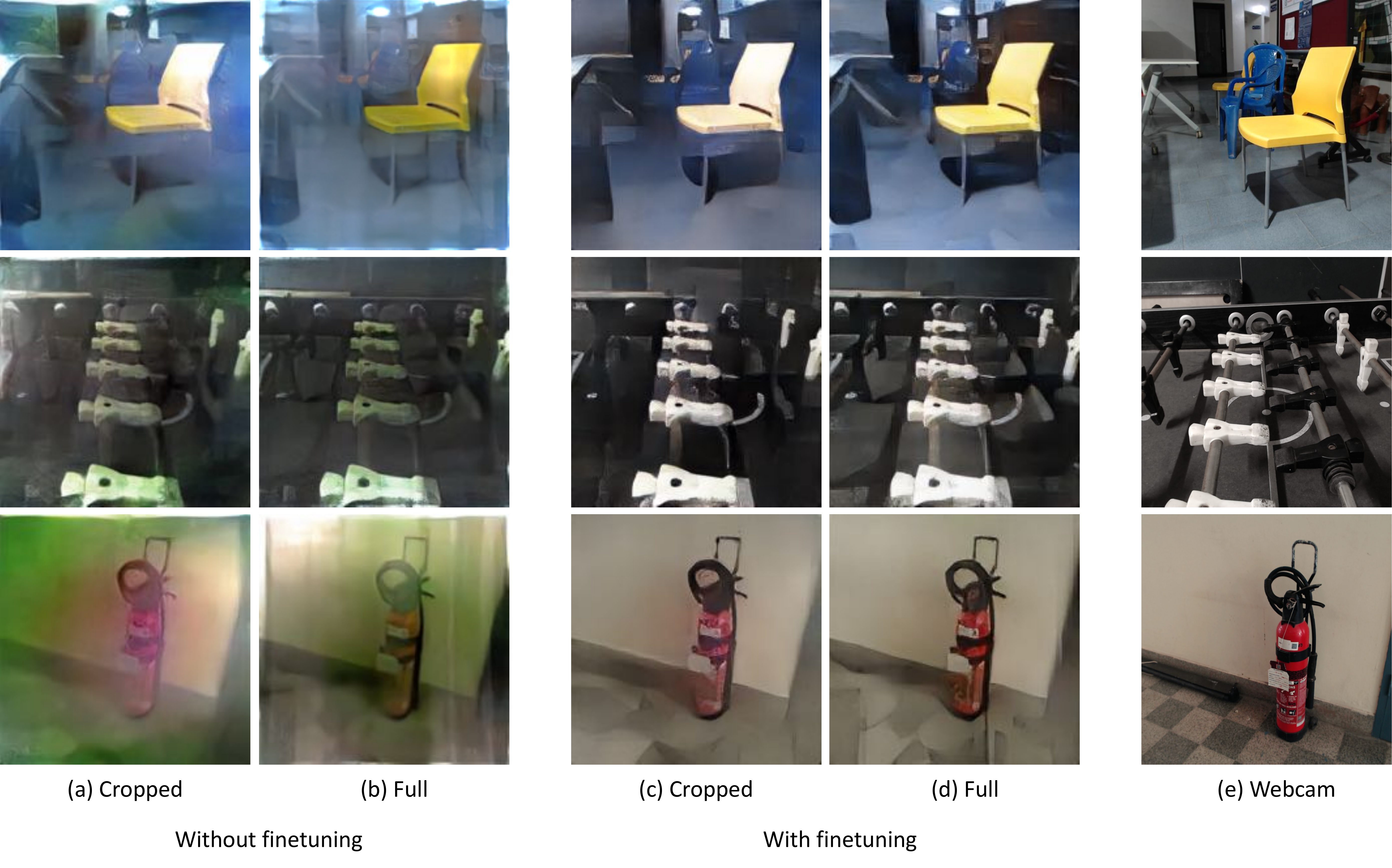}
    \begin{center}
    \caption{\textbf{Cropped measurements for Unconstrained Indoor Scenes}. We can observe that FlatNet-gen finetuned on unconstrained scenes provides reasonable reconstruction quality even for cropped measurements}
    \label{fig:finetune_crop}
    \end{center}
\end{figure*}

It is interesting to observe the effectiveness of the finetuned FlatNet for cropped unconstrained indoor scenes. In Figure \ref{fig:finetune_crop}, we provide visual comparison for the reconstructions from cropped measurement and full measurement along with the webcam capture. We show result for crop sizes of $990\times 1254$.  It should be noted that in an unconstrained setup, there may be large signals (due to bright objects) outside the field of view described by the CRA which would result in strong line artifacts in the reconstructions produced by model without finetuning.


\section{Effect of Bright Object}\label{bright}
For a highly multiplexed lensless imager,
every pixel receives light from every point in the scene. Hence, if there is any really bright object (like a highly reflective object or a lamp) in the scene, the light from the object can dominate the pixel intensities and result in severe reconstruction artifacts on the dimmer objects. We show that, using FlatNet, the artifacts are minimized resulting in a higher quality reconstruction of the scene.

\begin{figure*}[!ht]
\label{fig:dyn}
\centering
\includegraphics[scale=0.5]{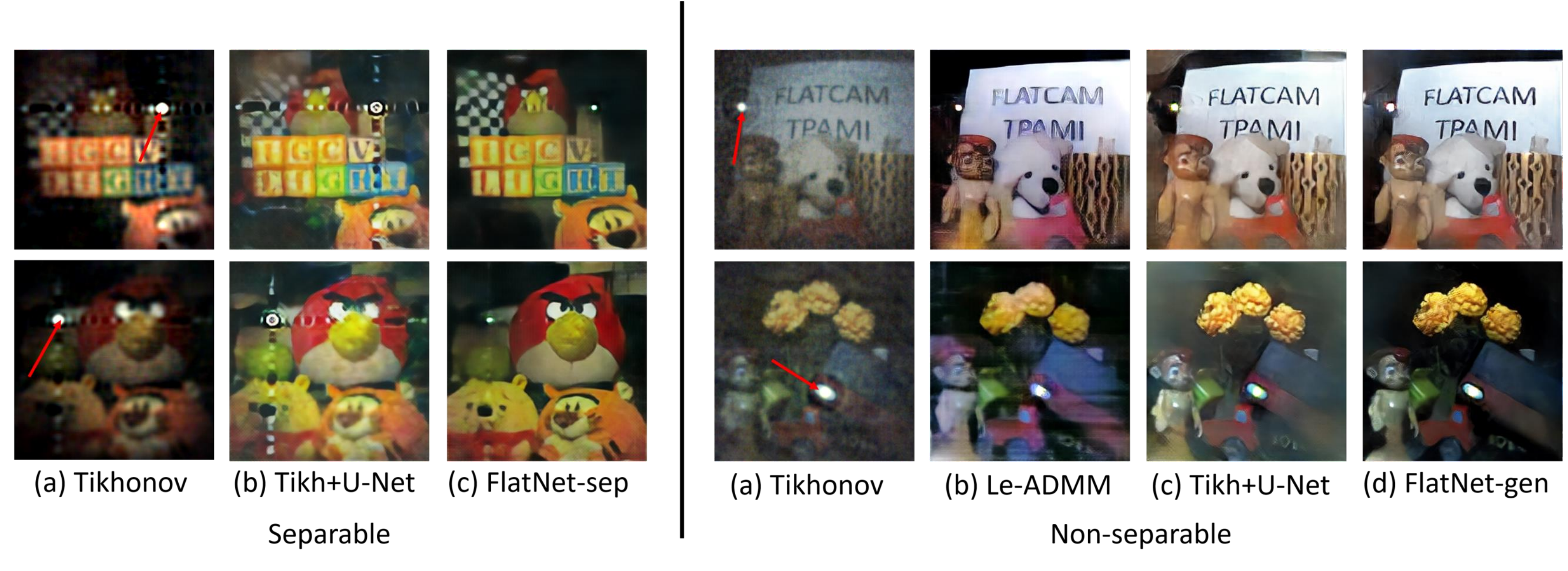}
\begin {center}
    \caption{\textbf{Reconstruction of scenes with bright objects (LED) using FlatCam and PhlatCam}. Artifacts occuring in Tikhonov reconstructions are amplified by Tikh+U-Net reconstruction. While Le-ADMM performs slightly better than Tikh+U-Net for PhlatCam, it is outperformed by FlatNet-gen}
    \label{fig:brightlight_comb}
    \end{center}
\end{figure*}

We show the bright object problem by introducing an
LED into the scene. Figure \ref{fig:brightlight_comb} shows the reconstruction for FlatCam\cite{asif2017flatcam} and PhlatCam\cite{boominathan2020phlatcam}. We can observe that FlatNet-sep and FlatNet-gen reconstructions have significantly fewer artifacts than other traditional and learning based approaches.

\ifpeerreview \else
\fi

\bibliographystyle{IEEEtran}